\documentclass[a4paper,12pt]{article}
\pdfoutput=1
\usepackage{amssymb}
\usepackage{amsmath}
\usepackage{amsfonts}
\usepackage{mathtools}
\usepackage{slashed}
\usepackage{color}
\usepackage{indentfirst}
\usepackage{graphicx}
\usepackage{bbm}
\usepackage{csquotes} 
\usepackage{caption}
\usepackage{cite}
\usepackage{here}
\usepackage{comment}
\usepackage{listings}
\usepackage[italicdiff]{physics}
\numberwithin{equation}{section}

\usepackage[unicode]{hyperref}
\usepackage{xcolor}
\hypersetup{
    colorlinks=false,
    citebordercolor=lime,
    linkbordercolor=pink,
}

\makeatletter
\newcommand*\rel@kern[1]{\kern#1\dimexpr\macc@kerna}
\newcommand*\widebar[1]{%
  \begingroup
  \def\mathaccent##1##2{%
    \rel@kern{0.8}%
    \overline{\rel@kern{-0.8}\macc@nucleus\rel@kern{0.2}}%
    \rel@kern{-0.2}%
  }%
  \macc@depth\@ne
  \let\math@bgroup\@empty \let\math@egroup\macc@set@skewchar
  \mathsurround\z@ \frozen@everymath{\mathgroup\macc@group\relax}%
  \macc@set@skewchar\relax
  \let\mathaccentV\macc@nested@a
  \macc@nested@a\relax111{#1}%
  \endgroup
}
\makeatother

\renewcommand{\thefootnote}{\fnsymbol{footnote}}

\begin{document}

\title{
\begin{flushright}
\begin{minipage}{0.25\linewidth}
\normalsize
KEK-TH-2421 \\
KYUSHU-HET-240 \\*[50pt]
\end{minipage}
\end{flushright}
{\Large \bf 
Spacetime evolution during moduli stabilization 
in radiation dominated era beyond 4D effective theory
\\*[20pt]}}

\author{
Hajime~Otsuka$^{a}$\footnote{
E-mail address: otsuka.hajime@phys.kyushu-u.ac.jp
}
\ and\
Yutaka~Sakamura$^{b,c}$\footnote{
E-mail address: sakamura@post.kek.jp
}\\*[20pt]
$^a${\it \normalsize
Department of Physics, Kyushu University,}\\ 
{\it \normalsize
744 Motooka, Nishi-ku, Fukuoka, 
819-0395, Japan}\\
$^b${\it \normalsize 
KEK Theory Center, Institute of Particle and Nuclear Studies, KEK,}\\
{\it \normalsize 1-1 Oho, Tsukuba, Ibaraki 305-0801, Japan}\\
$^c${\it \normalsize 
Graduate University for Advanced Studies (Sokendai),}\\
{\it \normalsize 1-1 Oho, Tsukuba, Ibaraki 305-0801, Japan.}
}
\maketitle

\date{
\centerline{\small \bf Abstract}
\begin{minipage}{0.9\linewidth}
\medskip 
\medskip 
\small
We investigate the time evolution of the background spacetime during the moduli stabilization process, 
which is assumed to occur in the radiation dominated era. 
The setup is basically the Salam-Sezgin model, but we add 
a potential term for the dilaton in order to stabilize the moduli completely. 
We numerically solve the higher-dimensional background field equations, including a case that the stabilization process cannot be described within the 4D effective theory. 
In contrast to the conventional 4D effective theory analysis, we find that when the mass scale of the stabilization is larger 
than the compactification scale, the radiation contribution to the total energy density 
remains to be non-negligible for a much longer time than the stabilization time scale. 
As a result, the non-compact 3D space expands slower than the matter dominated universe. 
We also find the equation of state for the radiation~$w_{\rm rad}$ remains to be smaller than 1/3 for a long time, 
which indicates that the radiation still feels the extra dimensions for a while even after the moduli are stabilized. 
\end{minipage}
}

\renewcommand{\thefootnote}{\arabic{footnote}}
\setcounter{footnote}{0}
\thispagestyle{empty}
\clearpage
\addtocounter{page}{-1}

\tableofcontents

\section{Introduction}

Current and future gravitational wave observations will give impacts on several stages of the cosmological history of our universe. 
There is a possibility to detect signatures of a fundamental theory of particle physics and cosmology, e.g., a higher-dimensional gravity arising from string theory, 
from the spectrum of the gravitational waves~\cite{Andriot:2017oaz,Yu:2019jlb} 
since their spectrum depends on the spacetime in which the gravitational waves have propagated. 
This direction of the search for the extra dimensions is complementary to the collider experiments 
that have searched for the Kaluza-Klein (KK) modes. 
In the early universe, the three-dimensional (3D) space was much smaller, and we expect that 
various effects due to the existence of the extra dimensions were larger than those in the current universe. 
Hence, in order to extract some signatures of the extra dimensions from the gravitational wave spectrum, 
we need to know the detailed time evolution of the background spacetime at early times. 

The stabilization of the moduli to some finite values is an important process to realize the standard 
four-dimensional (4D) cosmology at late times. 
In this paper, we investigate the time evolution of the space during the moduli stabilization, 
which is assumed to occur in the radiation dominated era. 
When the moduli stabilization can be described within the 4D effective theory, 
it was well known that the oscillation of the moduli around the potential minimum rapidly dominates the energy density 
and the 3D space expands as the matter dominated universe during the oscillation~\cite{Coughlan:1983ci,Goncharov:1984qm,Ellis:1986zt,Banks:1993en,deCarlos:1993wie}. 
However, in such a case, we must restrict the parameter space 
so that the typical mass scale of the stabilization~$m$ is lower than the compactification scale~$m_{\rm KK}$, 
in order for the 4D effective theory analysis to be valid. 
The purpose of this paper is to extend the analysis to a more general choice of the parameters, 
including cases in which $m\gtrsim m_{\rm KK}$. 
We numerically pursue the spacetime evolution by solving the higher-dimensional field equations 
including the radiation that fills in the bulk, instead of analyzing the 4D effective theory.

As a setup, we consider a gauged six-dimensional (6D) supergravity \cite{Nishino:1984gk,Randjbar-Daemi:1985tdc} 
compactified on a sphere~$S^2$. 
Since the 6D spacetime allows the background tensor fields such as fluxes on the internal space, 
the 6D gauged supergravity has a rich structure, such as the existence of chiral fermions\cite{Witten:1983ux,Witten:1984dg}, the Green-Schwarz anomaly cancellation\cite{Green:1984bx}, and a self-tuning of the four-dimensional vacuum energy\cite{Aghababaie:2003wz,Burgess:2004kd,Garriga:2004tq,Burgess:2004dh}. 
It is well known that a static solution is realized by balancing between the $S^2$ curvature and the $U(1)$ magnetic flux 
in a specific model, which is called the Salam-Sezgin model \cite{Salam:1984cj}. It was generalized to incorporate codimension-two branes \cite{Aghababaie:2003wz,Burgess:2004kd,Garriga:2004tq,Burgess:2004dh,Gibbons:2003di,Aghababaie:2003ar,Lee:2005az}. 
The time-dependent extension of the Salam-Sezgin background solution was discussed in Ref. \cite{Maeda:1984gq} 
in the vacuum and the radiation dominated situations. 
They obtained a solution that asymptotically approaches the 4D radiation dominated universe. 
In their analysis, however, the equation of state is set to a constant by hand. 
Besides, there remains a modulus that is not stabilized in their setup. 
In this work, we treat the pressure and the energy density independently, and pursue their time evolutions 
in the presence of a moduli-stabilization potential that completely fixes the moduli.

The paper is organized as follows. 
In Sec. \ref{sec:2}, we briefly review the Salam-Sezgin model \cite{Salam:1984cj} 
and its time-dependent extension: the Nishino-Maeda model \cite{Maeda:1984gq}. 
In Sec. \ref{sec:3}, we derive the expressions for various thermodynamic quantities in the five-dimensional (5D) space 
and the moduli potential by introducing an additional dilaton potential. 
In Sec. \ref{sec:3_3}, we numerically solve the background field equations including the radiation contribution, 
and see the time evolutions of the background space and the radiation contribution to the total energy density. 
Finally, we conclude this paper in Sec. \ref{sec:con}.

\section{6D supergravity compactified on a sphere}
\label{sec:2}

In this section, we briefly review the known background solutions of 6D supergravity compactified on a sphere.

\subsection{Setup}
\label{subsec:Setup}

We focus on a bosonic part of the gauged 6D ${\cal N}=(1,0)$ supergravity \cite{Nishino:1984gk,Randjbar-Daemi:1985tdc}, following the convention of Ref. \cite{Aghababaie:2002be}. 
The bosonic fields of our interest consist of a gravity multiplet including the metric ($g_{MN}$), 
the self-dual antisymmetric Kalb-Ramond field ($B_{MN}$), the dilaton ($\sigma$), and a $U(1)$ gauge field ($A_M$). 
The indices~$M,N=0,1,2,\cdots,5$ are the 6D coordinate indices. 
The 6D gravitational anomalies are canceled by the introduction of hypermultiplets or additional vector multiplets, 
but we assume that they do not affect the dynamics of background geometry at the moment. 
Then, the 6D effective action is given by
\begin{align}
\begin{split}
        S &= \int d^6 x \sqrt{-g^{(6)}}
                \biggl[-\frac{1}{2}R -\frac{1}{2}g^{MN}\partial_M \sigma \partial_N \sigma -\frac{g^2e^{\sigma}}{4}F^{MN}F_{MN}
                -\frac{g^4e^{2\sigma}}{12}G_{MNP}G^{MNP}-V(\sigma)\biggl],
\label{eq:6DSeff}          
\end{split}
\end{align}
where $g^{(6)}\equiv \det(g_{MN})$ is the determinant of the 6D metric, 
$F_{MN}\equiv\partial_M A_N -\partial_N A_M$ denotes the $U(1)$ gauge field strength with the gauge coupling $g$, and 
the three-form field strength for $B_{MN}$ is given by $G_{MNP}$ containing the Chern-Simons term: $G_{MNP}\equiv \partial_{[M}B_{NP]}+F_{[MN}A_{P]}$. 
Note that the three-form field strength $G_{MNP}$ including the Chern-Simons contribution 
is constructed in a $U(1)$ invariant way: 
\begin{align}
    \delta A_M = \partial_M \lambda, \quad
    \delta B_{MN} = -\lambda F_{MN}.
\end{align}
Here, we adopt the metric signature $(-,+,...,+)$ and the unit of the 6D gravitational constant $\kappa_{6}^2=8\pi G_6=1$ throughout this paper. 
We follow the Weinberg's curvature conventions for the Riemann curvature tensor as well as the covariant derivative of the gauge potential. 
The scalar potential is given by
\begin{align}
    V(\sigma) = 2e^{-\sigma},
    \label{eq:SSscalar}
\end{align}
which is induced as a consequence of the gauging of the $U(1)_R$ symmetry.

The equations of motions for $\{\sigma, B_{MN}, F_{MN}, g_{MN}\}$ are summarized as follows: 
\begin{align}
    \begin{split}
        &\frac{1}{\sqrt{-g^{(6)}}}\partial_M (\sqrt{-g^{(6)}}\partial^M \sigma)-\frac{g^2e^{\sigma}}{4}F^{MN}F_{MN} 
-\frac{g^4e^{2 \sigma}}{6}G_{MNP}G^{MNP} 
         -\partial_{\sigma}V(\sigma)=0,
        \\
        &\partial_M \left(\sqrt{-g^{(6)}}e^{2\sigma}G^{MNP}\right)=0,
        \\
        &\partial_M \left(\sqrt{-g^{(6)}}e^{\sigma}F^{MN}\right)+g^2e^{2\sigma}G^{MNP}F_{MP}=0,
        \\
        &R_{MN}-\frac{1}{2}g_{MN}R-T_{MN} = 0. 
    \end{split} \label{EoM}
\end{align}
The last equation is the Einstein equation. 
The explicit form of the energy-momentum tensor $T_{MN}$ is provided by
\begin{align}
\begin{split}
        T_{MN} =&\frac{2}{\sqrt{-g}}\frac{\delta S_{\rm matter}}{\delta g^{MN}}
    \\
       =&-\partial_M\sigma \partial_N \sigma
    +\frac{1}{2}g_{MN}\partial_L\sigma \partial^L \sigma
    -g^2e^{\sigma}F_{MC}F_{N}^{\,\,C}
    +\frac{g^2e^{\sigma}}{4}g_{MN}F^{AB}F_{AB}
    \\
    &-\frac{g^4 e^{2\sigma}}{2}G_{MBC}G_N^{\,\,BC}
    +\frac{g^4 e^{2\sigma}}{12}g_{MN}G_{ABC}G^{ABC}
    +g_{MN}V(\sigma).
\end{split}
\end{align}

\subsection{Salam-Sezgin model}
\label{subsec:SS}

We first review the famous Salam-Sezgin model \cite{Salam:1984cj} as a solution of 6D SUGRA on the sphere. 
The background fields are chosen as
\begin{align}
\begin{split}
    g_{MN} &=
    \begin{pmatrix}
    \eta_{\mu\nu} & 0\\
    0             & g_{mn}(y)
    \end{pmatrix}
    ,
    \\
    \sigma &= \mbox{(constant)},
    \\
    B_{MN} &= 0,
    \\
    F_{MN} &=
    \begin{pmatrix}
    0 & 0\\
    0 & F_{mn}(y)
    \end{pmatrix}
    ,   
\end{split} \label{SalamSezginBG}
\end{align}
where $\eta_{\mu\nu}$ ($\mu,\nu=0,1,2,3$) is the 4D Minkowski metric, and $y=(x^4,x^5)$ denotes the coordinates of the compact space. 
The background metric of $S^2$ is 
\begin{align}
\begin{split}
    g_{mn}(y) &= b^2r^2(d\theta^2 +\sin^2 \theta d\phi^2),
\end{split}    
\end{align}
where $m,n=4,5$, and $r$ is some fixed length scale. 
In this coordinate $\{x^4,x^5\}=\{\theta, \phi\}$, the radius of $S^2$ is given by $br$, 
and the lightest fluctuation mode of $b$ corresponds to the radion from the 4D point of view. 
In the following, we choose $r$ as the 6D Planck length, i.e., $r=1$. 
Here, we assumed that $B_{MN}$ does not have a nontrivial background, but 
the $U(1)$ gauge field~$A_M$ has the monopole background allowed by $S^2$ geometry. 
\begin{align}
    F_{mn}(y) &= f\epsilon_{mn}\equiv \frac{n}{2g b^2}\epsilon_{mn}, \;\;\;\;\; \left(n\in{\mathbb Z}\right)
\end{align}
where 
\begin{align}
 \epsilon_{45} &= -\epsilon_{54} = \sqrt{\det(g_{mn})} = b^2\sin\theta, \\
 \epsilon_{44} &= \epsilon_{55} = 0. 
\end{align}
This satisfies the equation of motion and the Dirac quantization condition, 
\begin{align}
    g\int_{S^2}d^2y\; F_{45} &= g\int_0^{2\pi}d\phi\int_0^\pi d\theta\;\frac{n}{2g}\sin\theta = 2\pi n.
\label{eq:Dirac}
\end{align}
The monopole charge $n$ is fixed to satisfy the equations of motion for the dilaton $\sigma$ and 
the internal metric $g_{mn}$:
\begin{align}
\begin{split}
    \frac{g^2e^{\sigma}}{4}F^{mn}F_{mn}&=-\partial_\sigma V(\sigma)=V(\sigma),
    \\
    R_{mn} &=- g^2e^{\sigma}F_{mp}F_n^{\;\,p}= - g^2e^{\sigma}f^2g_{mn},
\end{split}    
\end{align}
which are rewritten as
\begin{align}
\begin{split}
    e^{-\sigma} &= \frac{n^2}{4}b^{-2},
    \\
    e^{-2\sigma} &= \frac{n^2}{16}b^{-4},
\end{split}    
\end{align}
respectively. 
We have used the relation: $R_{mn} = b^{-2}g_{mn}$. 
Hence, these equations are consistent with each other iff the monopole charge is 
chosen as
\begin{align}
    n^2 =1,
\end{align}
as stated in Ref. \cite{Salam:1984cj}.
Note that the scalar potential (\ref{eq:SSscalar}) only stabilizes a linear combination 
of the dilaton~$\sigma$ and the size of $S^2$ (radion)~$b$. 
Namely, it only relates these fields as
\begin{align}
e^{\sigma} = 4 b^2,
\label{eq:SSvev}
\end{align}
and does not fix their individual values.

\subsection{Nishino-Maeda model}
\label{subsec:NM}

We next discuss time-dependent solutions in the Salam-Sezgin setup~\cite{Maeda:1984gq,Maeda:1985es}. 
We allow the time-dependence of the background, and modify (\ref{SalamSezginBG}) as 
\begin{align}
\begin{split}
    g_{MN} &=
    \begin{pmatrix}
    g_{\mu\nu}(t) & 0\\
    0             & g_{mn}(t,y)
    \end{pmatrix}
    ,
    \\
    \sigma &= \sigma(t),
    \\
    B_{MN} &= 0,
    \\
    F_{MN} &=
    \begin{pmatrix}
    0 & 0\\
    0 & F_{mn}(t,y)
    \end{pmatrix}
    ,    
\end{split} \label{NM:BGansatz}
\end{align}
where the background metric is assumed as 
\begin{align}
    ds^2 &= g_{MN}dx^M dx^N \nonumber\\
    &=-dt^2 +a(t)^2\left\{(dx^1)^2+(dx^2)^2+(dx^3)^2 \right\}
    +b(t)^2 (d\theta^2 +\sin^2 \theta d\phi^2),
\end{align}
with the scale factors $\{a(t), b(t) \}$, and the $U(1)$ magnetic field takes the value:
\begin{align}
    F_{mn}(t,y) &= f\epsilon_{mn}=\frac{n}{2g b^2(t)}\epsilon_{mn}, \;\;\;\;\; (n=\pm 1) \label{NM:BGansatz:3}
\end{align}
satisfying the equation of motion for $F_{mn}$ and the Dirac quantization condition (\ref{eq:Dirac}). 
In the following analysis, we set the monopole charge $n=1$. 
There are two independent classes of solutions depending on the initial conditions.
\begin{enumerate}
\item Static solution:
\begin{align}
\begin{split}
            a &= a_0,
         \\
        b &= b_0,
          \\
        \sigma &= \sigma_0, 
\end{split}
\end{align}
where $\{a_0, b_0, \sigma_0\}$ are constants and this corresponds to the Salam-Sezgin solution.
\item Time-dependent solution:
\begin{align}
\begin{split}
            a &= a_0 (t - t_0)^{\frac{9\pm 4\sqrt{3}}{33}},
         \\
        b &= b_0 (t - t_0)^{\frac{1\mp  2\sqrt{3}}{11}},
          \\
        \sigma &= \sigma_0 +\frac{2(1\mp 2\sqrt{3})}{11}\ln (t - t_0),
\end{split}
\end{align}
where $\{a_0, b_0, \sigma_0,t_0\}$ are constants. 
Noting that $(9+ 4\sqrt{3})/33\simeq 0.48$, one of these solutions has a similar expansion property to the 4D radiation dominated one. 
But the compact extra space continues to shrink in that case. 
\end{enumerate}

In addition to the above solutions, the authors of Refs.~\cite{Maeda:1984gq,Maeda:1985es} also introduce 
the radiation to the original action,
\begin{align}
    S\rightarrow S+S_{\rm rad}. 
\end{align}
The radiation consists of the gravitino and/or the matter fields that are 
required by the cancellation of the 6D gravitational anomalies. 
Such fields are supposed to be in the thermal equilibrium. 
They assume that the energy-momentum tensor for the radiation has the form of
\begin{align}
 \left(T_{\rm rad}\right)_M^{\;\;\;N} &= \begin{pmatrix} \rho_{\rm rad} & & \\ 
 & -p_{{\rm rad},3}\mbox{\boldmath $1_3$} & \\ & & -p_{{\rm rad},2}\mbox{\boldmath $1_2$} \end{pmatrix}. 
 \label{diag:T_rad}
\end{align}
They also assume that the equations of state for the radiation are constants. 
For example, when 
\begin{align}
 p_{\rm rad,3} &= \frac{\rho_{\rm rad}}{3}, \;\;\;\;\;
 p_{\rm rad,2} = 0, \label{anisotropic_p}
\end{align}
they found that the background spacetime asymptotically approaches the 4D radiation dominated universe at late times. 
\begin{align}
 a(t) &\to a_0 t^{1/2}, \;\;\;\;\;
 b(t) \to b_0, \;\;\;\;\;
 \sigma(t) \to \sigma_0, \nonumber\\
 F_{\mu\nu} & = F_{\mu m} = 0, \;\;\;\;\;
 F_{mn}(t) = \frac{1}{2gb^2(t)}\epsilon_{mn}, \;\;\;\;\;
 B_{MN} = 0,  \label{NM:sol}
\end{align}
where $a_0$, $b_0$ and $\sigma_0$ are constants. 
However, the equations of state~$w_{\rm rad,3}\equiv p_{\rm rad,3}/\rho_{\rm rad}$ and $w_{\rm rad,2}\equiv p_{\rm rad,2}/\rho_{\rm rad}$ 
should be determined by the dynamics, and thus change with time. 
Thus, in the following sections, we will numerically solve the evolution equations for the background, 
and see how the background spacetime and $w_{\rm rad,3}$ evolve with time.

\section{Thermodynamic quantities and moduli potential}
\label{sec:3}

In this section, we derive the expressions for the thermodynamic quantities for the 6D radiation 
and the scalar potential for the moduli.

\subsection{Thermodynamic quantities}
\label{sec:3_1}

The physical volume of the compact space~$S^2$ is given by 
\begin{align}
 {\cal V}_2 &= 4\pi b^2. 
\end{align}
Since the comoving volume for the 3D space is 
\begin{align}
  {\cal V}_3 &= a^3, 
 \end{align}
the 5D comoving volume is given by
\begin{align}
    \begin{split}
{\cal V}_5 = {\cal V}_3{\cal V}_2 = 4\pi a^3 b^2.
    \end{split}
\end{align}

The dispersion relation of a 6D massless particle on this background:
\begin{align}
    \begin{split}
P_MP^M = -p_0^2 +\frac{1}{a^2}\vec{p}^{\;2}+\frac{1}{b^2}p_\theta^2+\frac{1}{b^2\sin^2\theta}p_\phi^2 = 0
    \end{split}
\end{align}
implies that the energy of the particle with the 3D momentum $\vec{p}=(p_1,p_2,p_3)$ and the angular momentum $l$ on $S^2$ is 
\begin{align}
    \begin{split}
{\cal E}_{p,l} \equiv p_0 = \sqrt{\frac{p^2}{a^2}+\frac{l(l+1)}{b^2}},
    \end{split}
    \label{E_pl}
\end{align}
where $p\equiv\sqrt{\vec{p}^{\;2}}$. 
Since each one-particle state is specified by $\vec{p}$, $l$ and the `magnetic quantum number'~$m=-l,\cdots,l$, 
we have $(2l+1)$ degenerate energy eigenstates for each $\vec{p}$ and $l$.

The grand potential is then expressed as
\begin{align}
 J(\beta,\mu,{\cal V}_3,{\cal V}_2) &= \pm\sum_{l=0}^\infty\frac{g_{\rm dof}(2l+1)}{2\pi^2\beta}
  \int_0^\infty dp\;p^2\ln\left(1\mp e^{-\beta({\cal E}_{p,l}-\mu)}\right),
  \label{expr:J}
\end{align}
where $g_{\rm dof}$ denotes the degrees of freedom for the 6D relativistic particles, 
$\beta$ is the inverse temperature, and $\mu$ is the chemical potential. 
The upper sign in the logarithm corresponds to the case of bosons, while the lower is that of fermions. 

From (\ref{expr:J}), various thermodynamic quantities are calculated as follows. 
\begin{itemize}
\item
Energy density: 
\begin{align}
 \rho_{\rm rad} &= \frac{1}{{\cal V}_5}\left(\partial_\beta-\frac{\mu}{\beta}\partial_\mu\right)(\beta J) 
 = \sum_{l=0}^\infty\frac{g_{\rm dof}(2l+1)}{2\pi^2{\cal V}_5}
 \int_0^\infty dp\;\frac{p^2{\cal E}_{p,l}}{e^{\beta({\cal E}_{p,l}-\mu)}\mp 1} \nonumber\\
 &= \sum_{l=0}^\infty \frac{g_{\rm dof}(2l+1)}{2\pi^2\beta^4{\cal V}_2}\int_0^\infty dk\;
 \frac{k^2\sqrt{k^2+c_l^2}}{e^{\sqrt{k^2+c_l^2}-\beta\mu}\mp 1}. 
 \label{rho_rad:exact}
\end{align}
In the second line, we have rescaled the integration variable and the KK masses as
\begin{align}
 k &\equiv \frac{\beta}{a}p, \;\;\;\;\;
 c_l \equiv \beta\sqrt{\frac{4\pi l(l+1)}{{\cal V}_2}}. 
 \label{rescale}
\end{align}
\item
Pressures: \\
The pressure in the 3D non-compact space is given by
\begin{align}
 P_3 &= -\frac{\partial J}{\partial {\cal V}_3} = \sum_{l=0}^\infty\frac{g_{\rm dof}(2l+1)}{6\pi^2a^5}\int_0^\infty dp\;
 \frac{p^4}{{\cal E}_{p,l}(e^{\beta({\cal E}_{p,l}-\mu)}\mp 1)} \nonumber\\
 &= \sum_{l=0}^\infty \frac{g_{\rm dof}(2l+1)}{6\pi^2\beta^4}\int_0^\infty dk\;
 \frac{k^4}{\sqrt{k^2+c_l^2}(e^{\sqrt{k^2+c_l^2}-\beta\mu}\mp 1)}, 
 \label{P_3:exact}
\end{align}
and that in the $S^2$ compact space is given by
\begin{align}
 P_2 &= -\frac{\partial J}{\partial{\cal V}_2} 
 = \sum_{l=0}^\infty\frac{g_{\rm dof}l(l+1)(2l+1)}{\pi {\cal V}_2^2}\int_0^\infty dp\;
 \frac{p^2}{{\cal E}_{p,l}(e^{\beta({\cal E}_{p,l}-\mu)}\mp 1)} \nonumber\\
 &= \sum_{l=1}^\infty\frac{g_{\rm dof}l(l+1)(2l+1){\cal V}_3}{\pi {\cal V}_2^2\beta^2}
 \int_0^\infty dk\;\frac{k^2}{\sqrt{k^2+c_l^2}(e^{\sqrt{k^2+c_l^2}-\beta\mu}\mp 1)}. 
 \label{P_2:exact}
\end{align}
\item
Entropy:
\begin{align}
 {\cal S}_{\rm rad} &= \beta^2\frac{\partial J}{\partial\beta}
 = \sum_{l=0}^\infty\frac{g_{\rm dof}(2l+1)}{2\pi^2}\int_0^\infty dp\;p^2
 \left\{\mp\ln\left(1\mp e^{-\beta({\cal E}_{p,l}-\mu)}\right)
 +\frac{\beta({\cal E}_{p,l}-\mu)}{e^{\beta({\cal E}_{p,l}-\mu)}\mp 1}\right\} \nonumber\\
 &= \sum_{l=0}^\infty\frac{g_{\rm dof}(2l+1){\cal V}_3}{2\pi^2\beta^3}
 \int_0^\infty dk\;k^2\left\{\mp\ln\left(1\mp e^{-\sqrt{k^2+c_l^2}+\beta\mu}\right)
 +\frac{\sqrt{k^2+c_l^2}-\beta\mu}{e^{\sqrt{k^2+c_l^2}-\beta\mu}\mp 1}\right\}. 
 \label{cS:exact}
\end{align}
\end{itemize}

The equation~(\ref{expr:J}) is rewritten as
\begin{align}
 J(\beta,\mu,{\cal V}_3,{\cal V}_2)
 &= \mp\frac{g_{\rm dof}{\cal V}_3}{\pi^2\beta^4}{\rm Li}_4(\pm e^{\beta\mu})
 \pm\sum_{l=1}^\infty\frac{g_{\rm dof}(2l+1){\cal V}_3}{2\pi^2\beta^4}
 \int_0^\infty dk\;k^2\ln\left(1\mp e^{-\sqrt{k^2+c_l^2}+\beta\mu}\right), 
\end{align}
where ${\rm Li}_s(z)$ is the polylogarithmic function. 
The first term is the contribution of the $l=0$ state. 
In the following, we consider a situation in which $e^{-c_l+\beta\mu}\ll 1$ ($l\geq 1$). 
Then, the grand potential can be approximated as
\begin{align}
 J(\beta,\mu,{\cal V}_3,{\cal V}_2)
 &\simeq \mp\frac{g_{\rm dof}{\cal V}_3}{\pi^2\beta^4}{\rm Li}_4(\pm e^{\beta\mu})
 -\sum_{l=1}^\infty\frac{g_{\rm dof}(2l+1){\cal V}_3}{2\pi^2\beta^4}
 \int_0^\infty dk\;k^2 e^{-\sqrt{k^2+c_l^2}+\beta\mu} \nonumber\\
 &= -\frac{g_{\rm dof}{\cal V}_3}{2\pi^2\beta^4}\left\{
 \pm 2{\rm Li}_4(\pm e^{\beta\mu})+e^{\beta\mu}Q_1\left(\beta\sqrt{\frac{4\pi}{{\cal V}_2}}\right)\right\}, 
\end{align}
where 
\begin{align}
 Q_1(x) &\equiv \sum_{l=1}^\infty x^2 l(l+1)(2l+1)K_2\left(x\sqrt{l(l+1)}\right). 
 \label{def:Q_1}
\end{align}
Here $K_2(z)$ is the modified Bessel function of the second kind. 
Then the above quantities are approximated as 
\begin{align}
 \rho_{\rm rad} &\simeq \rho_{\rm rad}^{\rm ap} 
 \equiv \frac{g_{\rm dof}}{2\pi^2\beta^4 {\cal V}_2}
 \left\{\pm 6{\rm Li}_4(\pm e^{\beta\mu})+e^{\beta\mu}(3Q_1+Q_2)\right\}, \nonumber\\
 P_3 &\simeq P_3^{\rm ap} \equiv \frac{g_{\rm dof}}{2\pi^2\beta^4}
 \left(\pm 2{\rm Li}_4(\pm e^{\beta\mu})+e^{\beta\mu}Q_1\right), \nonumber\\
 P_2 &\simeq P_2^{\rm ap} \equiv \frac{g_{\rm dof}e^{\beta\mu}{\cal V}_3}{4\pi^2\beta^4 {\cal V}_2}Q_2, \nonumber\\
 {\cal S}_{\rm rad} &\simeq {\cal S}^{\rm ap}_{\rm rad} 
 \equiv \frac{g_{\rm dof}{\cal V}_3}{2\pi^2\beta^3}\left\{\pm 8{\rm Li}_4(\pm e^{\beta\mu})+e^{\beta\mu}(4Q_1+Q_2)
 -\beta\mu\left(\pm 2{\rm Li}_3(\pm e^{\beta\mu})+e^{\beta\mu}Q_1\right)\right\}, 
 \label{expr:rho_rad}
\end{align}
where the arguments of $Q_i$ ($i=1,2,3$) are $\beta\sqrt{4\pi/{\cal V}_2}$, and 
\begin{align}
 Q_2(x) &\equiv \sum_{l=1}^\infty x^3l^{3/2}(l+1)^{3/2}(2l+1)K_1\left( x\sqrt{l(l+1)}\right). 
 \label{def:Q_2}
\end{align}
Here we have used that
\begin{align}
 Q_1'(x) &= -\frac{Q_2(x)}{x}, 
\end{align}
which follows from the formula for the modified Bessel functions: 
\begin{align}
 K'_\nu(z) = -K_{\nu-1}(z)-\frac{\nu}{z}K_\nu(z). 
 \label{der:K_nu}
\end{align}

Note that $p_{\rm rad,3}$ and $p_{\rm rad,2}$ in (\ref{diag:T_rad}) should be identified as
\begin{align}
 p_{\rm rad,3} &= \frac{P_3}{{\cal V}_2} \simeq p_{\rm rad,3}^{\rm ap} \equiv \frac{P_3^{\rm ap}}{{\cal V}_2}, \nonumber\\
 p_{\rm rad,2} &= \frac{P_2}{{\cal V}_3} \simeq p_{\rm rad,2}^{\rm ap} \equiv \frac{P_2^{\rm ap}}{{\cal V}_3}.  \label{def:p23}
\end{align}
From (\ref{expr:rho_rad}), we find that~\footnote{
From (\ref{rho_rad:exact}), (\ref{P_3:exact}) and (\ref{P_2:exact}), the first relation also holds for the exact expressions: 
\begin{align}
 \rho_{\rm rad} &= 3p_{\rm rad,3}+2p_{\rm rad,2}. 
 \nonumber
\end{align}
}
\begin{align}
 \rho_{\rm rad}^{\rm ap} &= 3p_{\rm rad,3}^{\rm ap}+2p_{\rm rad,2}^{\rm ap}, \nonumber\\
 s^{\rm ap}_{\rm rad} &\equiv \frac{{\cal S}^{\rm ap}_{\rm rad}}{{\cal V}_5} 
 = \beta \left\{(4-\beta\mu)p_{\rm rad,3}^{\rm ap}+2p_{\rm rad,2}^{\rm ap}\right\}. 
 \label{rel:rho-p}
\end{align}

The (radiation) energy density for the 4D effective theory is
\begin{align}
 \rho_{\rm rad}^{\rm (4D)} &= \rho_{\rm rad}{\cal V}_2. 
\end{align}
In the case that the energy density is dominated by $\rho_{\rm rad}$ 
and $p_{\rm rad,3}^{\rm ap}\gg p_{\rm rad,2}^{\rm ap}$, we have the usual equation of state in the 3D space:
\begin{align}
 w_{\rm rad} \equiv \frac{P_3}{\rho_{\rm rad}^{\rm (4D)}} = \frac{p_{\rm rad,3}}{\rho_{\rm rad}} 
 \simeq \frac{p_{\rm rad,3}^{\rm ap}}{\rho_{\rm rad}^{\rm ap}} \simeq \frac{1}{3}. 
\end{align}
If the pressure is isotropic in the whole 5D space, i.e., $p_{\rm rad,3}^{\rm ap}\simeq p_{\rm rad,2}^{\rm ap}$, this ratio becomes
\begin{align}
 w_{\rm rad} &\simeq \frac{1}{5}. 
\end{align}
Namely, $w_{\rm rad}^{-1}$ measures the dimension of the space that the radiation feels. 

\begin{figure}[H]
  \begin{center}
    \includegraphics[scale=0.8]{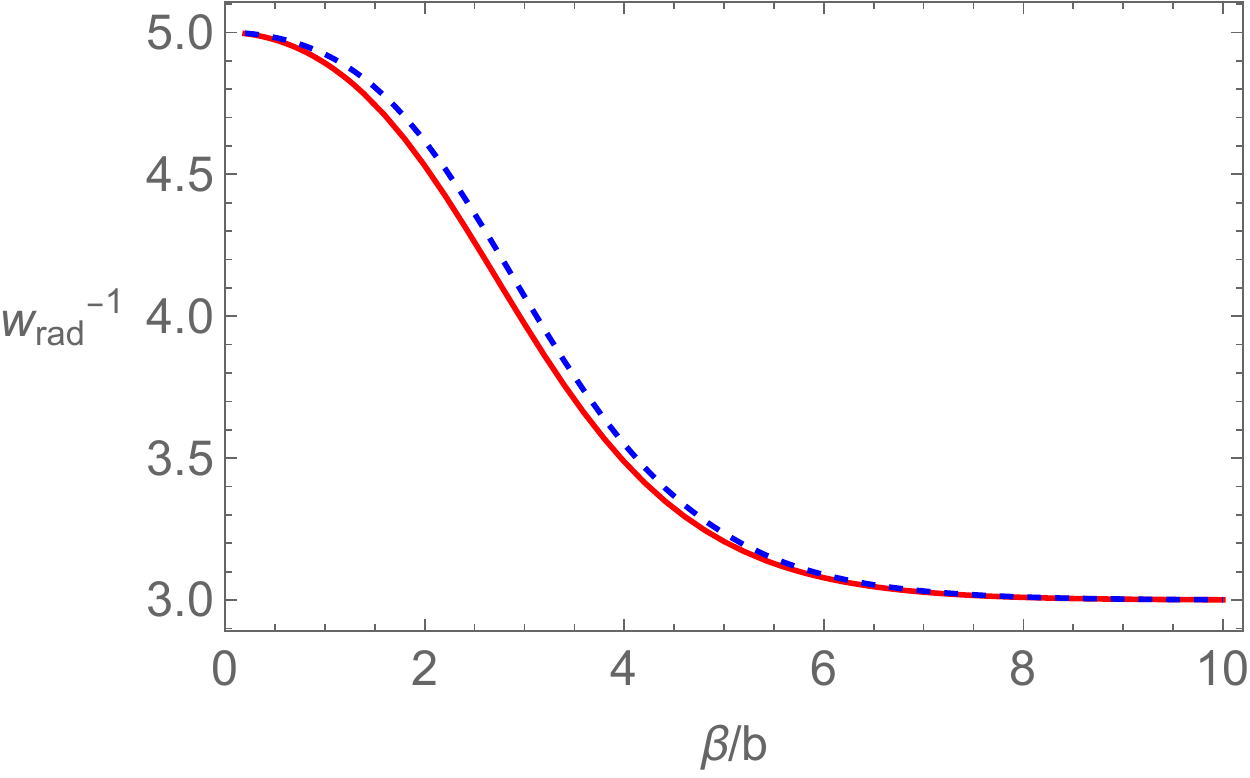} 
  \end{center}
\caption{The ratio~$w_{\rm rad}^{-1}$ in (\ref{expr:winv}). The red solid line corresponds to the bosonic case, while the blue dashed line corresponds to the fermionic case.}
    \label{fig:winv}
\end{figure}

Using  (\ref{expr:rho_rad}) and (\ref{rel:rho-p}), this ratio is expressed as
\begin{align}
 w_{\rm rad}^{-1} &\simeq 3+\frac{2p_{\rm rad,2}^{\rm ap}}{p_{\rm rad,3}}
 = 3+\frac{Q_2(x)}{\pm 2e^{-\beta\mu}{\rm Li}_4(\pm e^{\beta\mu})+Q_1(x)}, 
 \label{expr:winv}
\end{align}
where $x=\beta\sqrt{4\pi/{\cal V}_2}=\beta/b$. 
In the case that $\beta\mu\ll 1$ and negligible, $w_{\rm rad}^{-1}$ is a function of only $\beta/b$, 
which is plotted in Fig.~\ref{fig:winv}.\footnote{ 
We can numerically check that the approximate expression~(\ref{expr:winv}) well agrees with the exact expression 
used by (\ref{P_3:exact}) and (\ref{P_2:exact}) even for a small $\beta/b$. 
}
From this plot, we can see that the radiation feels almost all space dimensions when the extra space is large 
or the temperature is high ($\beta/b\ll 1$), 
but eventually feels only the non-compact 3D space at low temperature after the compact space is stabilized to a finite value ($\beta/b\gg 1$). 
However, if the compact space~$S^2$ continues to expand and the ratio~$\beta/b$ remains to be ${\cal O}(1)$, 
the value~$w_{\rm rad}^{-1}$ takes some intermediate value between 3 and 5. 

\begin{figure}[H]
  \begin{center}
    \includegraphics[scale=0.65]{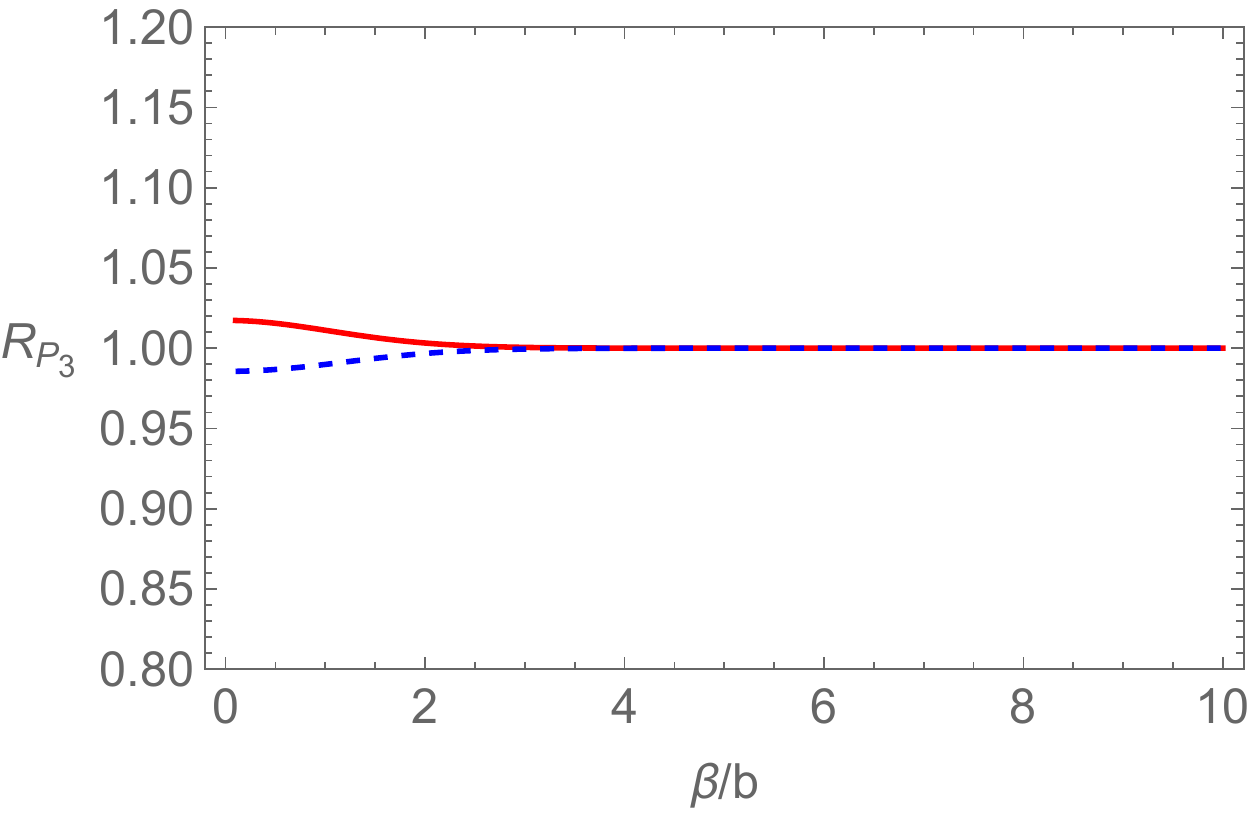} \;\;
    \includegraphics[scale=0.65]{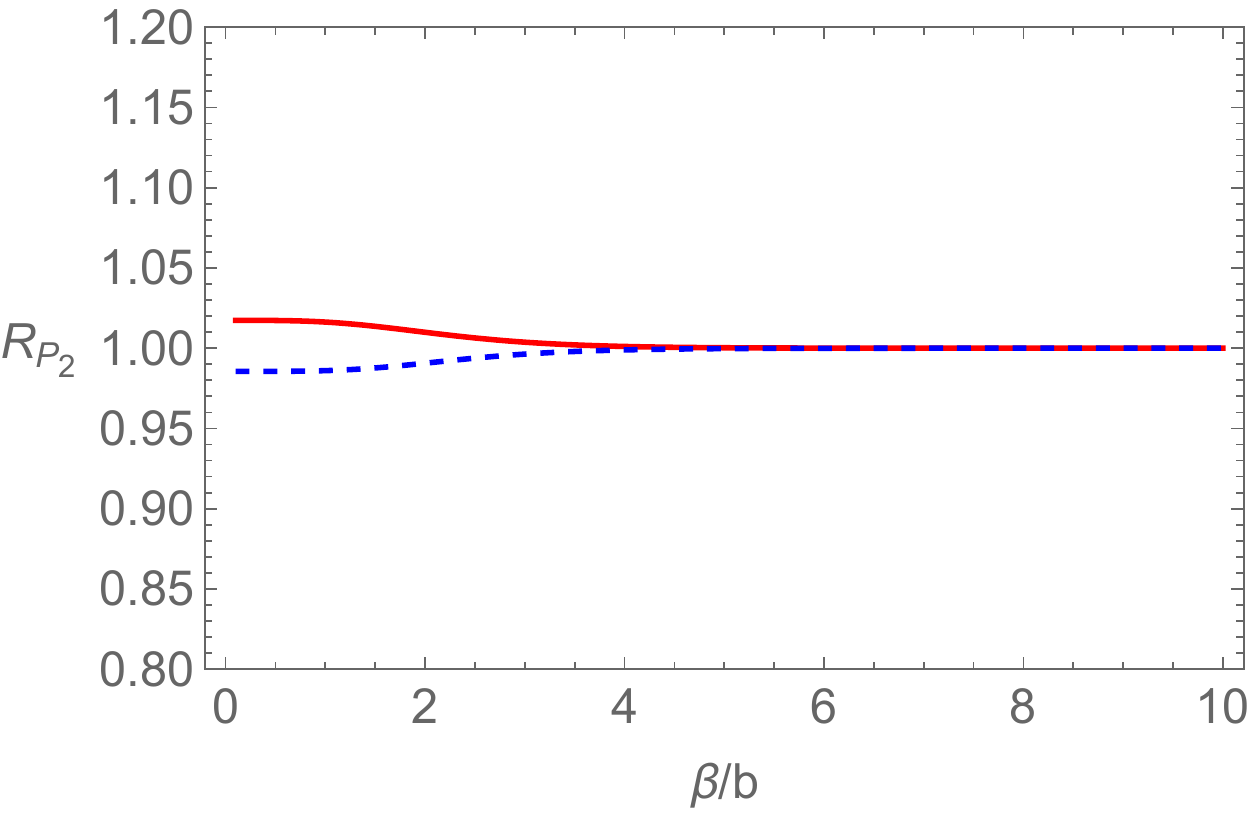}
  \end{center}
\caption{The ratios~${\cal R}_{P_3}$ (left plot) and ${\cal R}_{P_2}$ (right plot) in (\ref{expr:R_P}). 
The upper (red) solid lines correspond to the bosonic case, while the lower (blue) dashed line to the fermionic case. }
    \label{fig:R_P}
\end{figure}

In order to check the accuracy of the approximation in (\ref{expr:rho_rad}), 
we define the ratios: 
\begin{align}
 {\cal R}_{P_3} &\equiv \frac{P_3}{P_3^{\rm ap}}, \;\;\;\;\;
 {\cal R}_{P_2} \equiv \frac{P_2}{P_2^{\rm ap}}.  
 \label{def:R_P}
\end{align}
In the case that $\beta\mu\ll 1$ and negligible, we can see 
from (\ref{P_3:exact}), (\ref{P_2:exact}) and (\ref{expr:rho_rad}) 
that they are functions of only $\beta\sqrt{4\pi/{\cal V}_2}=\beta/b$. 
Namely, we have 
\begin{align}
 {\cal R}_{P_3}(x) \equiv& 
 \frac{1}{\pm 2{\rm Li}_4(\pm 1)+Q_1(x)}
 \sum_{l=0}^\infty\frac{2l+1}{3}\int_0^\infty dk\frac{k^4}{\sqrt{k^2+x^2l(l+1)}\left(e^{\sqrt{k^2+x^2l(l+1)}}\mp 1\right)}, \nonumber\\
 {\cal R}_{P_2}(x) \equiv& 
 \frac{x^2}{Q_2(x)}
 \sum_{l=1}^\infty l(l+1)(2l+1)\int_0^\infty dk\;
 \frac{k^2}{\sqrt{k^2+x^2l(l+1)}\left(e^{\sqrt{k^2+x^2l(l+1)}}\mp 1\right)}, 
 \label{expr:R_P}
\end{align}
where $x\equiv \beta/b$. 
Note that 
\begin{align}
 {\rm Li}_4(1) = \zeta(4), \;\;\;\;\;
 {\rm Li}_4(-1) = -\frac{7}{8}\zeta(4). 
\end{align}
Fig.~\ref{fig:R_P} shows the ratios in (\ref{def:R_P}) as functions of $\beta/b$. 
From these plots, we can see that the approximated quantities in (\ref{expr:rho_rad}) can be used 
for $\beta/b\gtrsim {\cal O}(1)$ with good accuracy. 

\subsection{Conservation law and entropy}
The energy-momentum conservation law is 
\begin{align}
 \nabla_M T^M_{\;\;N} \equiv \partial_M T^M_{\;\;N}+\Gamma^M_{\;\;ML}T^L_{\;\;N}-\Gamma^L_{\;\;MN}T^M_{\;\;L} = 0, 
 \label{T:conservation}
\end{align}
where 
\begin{align}
 T^t_{\;\;t} &= \frac{1}{2}\dot{\sigma}^2+\frac{e^\sigma}{8b^4}+2e^{-\sigma}+V_{\rm stab}(\sigma)+\rho_{\rm rad} \equiv \rho_{\rm tot}, 
 \nonumber\\
 T^i_{\;\;j} &= \delta^i_{\;\;j}\left\{-\frac{1}{2}\dot{\sigma}^2+\frac{e^\sigma}{8b^4}+2e^{-\sigma}+V_{\rm stab}(\sigma)-p_{\rm rad,3}\right\}
 \equiv -\delta^i_{\;\;j}p_{\rm tot,3}, \nonumber\\
 T^4_{\;\;4} &= T^5_{\;\;5} = -\frac{1}{2}\dot{\sigma}^2-\frac{e^\sigma}{8b^4}+2e^{-\sigma}+V_{\rm stab}(\sigma)-p_{\rm rad,2} 
 \equiv -p_{\rm tot,2}. 
 \label{Tcomp}
\end{align}
From the case of $N=t$ in (\ref{T:conservation}), we have~\footnote{
This is also obtained from the Einstein equations in (\ref{EOMs:abs}). 
}
\begin{align}
 \dot{\rho}_{\rm tot}+\frac{3\dot{a}}{a}\left(\rho_{\rm tot}+p_{\rm tot,3}\right)+\frac{2\dot{b}}{b}\left(\rho_{\rm tot}+p_{\rm tot,2}\right) = 0, 
 \label{EMconserve}
\end{align}
where the dot denotes the time derivative. 
The other components hold trivially, and do not provide any information. 
The conservation law~(\ref{EMconserve}) is rewritten as
\begin{align}
 \frac{d}{dt}\left(\rho_{\rm tot}{\cal V}_5\right) &= -p_{\rm tot,3}{\cal V}_2\frac{d{\cal V}_3}{dt}
 -p_{\rm tot,2}{\cal V}_3\frac{d{\cal V}_2}{dt}. 
\end{align}
Since $U_{\rm tot}\equiv \rho_{\rm tot}{\cal V}_5$ is the total energy in the 5D comoving volume and 
\begin{align}
 dU_{\rm tot} &= \frac{1}{\beta}d{\cal S}_{\rm tot}-P_3d{\cal V}_3-P_2d{\cal V}_2+\mu dN,  
\end{align}
where $N$ is the particle number, the total entropy in the 5D comoving volume~${\cal S}_{\rm tot}$ is conserved 
when we neglect the chemical potential~$\mu$. 
\begin{align}
 \dot{\cal S}_{\rm tot} &= 0. 
\end{align}

By using the dilaton equation of motion, the equation~(\ref{EMconserve}) is reduced to 
\begin{align}
 \dot{\rho}_{\rm rad}+\left(\frac{3\dot{a}}{a}+\frac{2\dot{b}}{b}\right)\rho_{\rm rad}
 +\frac{3\dot{a}}{a}p_{\rm rad,3}+\frac{2\dot{b}}{b}p_{\rm rad,2} &= 0.  \label{conserv:rad}
\end{align}
Thus, the conservation law holds only for the radiation. 
Namely, the radiation part of the entropy~${\cal S}_{\rm rad}$ is individually conserved 
when $\mu$ is negligible. 
\begin{align}
 \dot{\cal S}_{\rm rad} &= 0. 
 \label{conserve:cS_rad}
\end{align}

Plugging the expressions in (\ref{expr:rho_rad}) into (\ref{conserv:rad}), we obtain
\begin{align}
 &\frac{\dot{\beta}}{\beta}\left\{\pm 24{\rm Li}_4(\pm e^{\beta\mu})+e^{\beta\mu}\left(12Q_1+5Q_2+Q_3\right) \right. \nonumber\\
 &\hspace{5mm}\left. 
 -\beta\mu\left(\pm 6{\rm Li}_3(\pm e^{\beta\mu})+e^{\beta\mu}(3Q_1+Q_2)\right)\right\} \nonumber\\
 =&\frac{3\dot{a}}{a}\left\{\pm 8{\rm Li}_4(\pm e^{\beta\mu})+e^{\beta\mu}(4Q_1+Q_2)\right\}
 +\frac{\dot{b}}{b}e^{\beta\mu}(2Q_2+Q_3), \label{expr:dotbeta}
\end{align}
where
\begin{align}
 Q_3(x) &\equiv \sum_{l=1}^\infty x^4l^2(l+1)^2(2l+1)K_0\left(x\sqrt{l(l+1)}\right). 
 \label{def:Q_3}
\end{align}
We have used that
\begin{align}
 Q_2'(x) &= \frac{1}{x}\left\{2Q_2(x)-Q_3(x)\right\}, 
\end{align}
which is followed from the formula~(\ref{der:K_nu}). 
Eq.(\ref{expr:dotbeta}) is the evolution equation for the inverse temperature~$\beta$.

\subsection{Approximation of $\mbox{\boldmath $Q_i$ ($i=1,2,3$)}$}
\label{sec:3_15}
The definitions of the functions in (\ref{def:Q_1}), (\ref{def:Q_2}) and (\ref{def:Q_3}) involve infinite sums for $l$. 
In fact, we need to sum the summands up to large numbers of $l$ in order to reproduce the correct values of $Q_i(x)$ $(i=1,2,3)$ 
for $x\ll 1$ with high accuracy. 
However, the contributions from large $l$ become less important as $x$ increases. 
Thus, we approximate them by the following functions: 
\begin{align}
 Q_1^{\rm ap}(x) &\equiv \begin{cases} q_1(x) & (0< x < 1.0) \\
 \tilde{Q}_1(x;11) & (1.0\leq x < 2.0) \\
 \tilde{Q}_1(x;5) & (2.0\leq x < 5.0) \\
 \tilde{Q}_1(x;2) & (x\geq 5.0) \end{cases}, \nonumber\\
 Q_2^{\rm ap}(x) &\equiv \begin{cases} q_2(x) & (0< x < 2.0) \\
 \tilde{Q}_2(x;6) & (2.0\leq x < 5.0) \\
 \tilde{Q}_2(x;2) & (x\geq 5.0) \end{cases}, \nonumber\\
 Q_3^{\rm ap}(x) &\equiv \begin{cases} q_3(x) & (0< x < 3.0) \\
 \tilde{Q}_3(x;5) & (3.0 \leq x < 6.0) \\
 \tilde{Q}_3(x;2) & (x\geq 6.0) \end{cases}, 
 \label{apQs}
\end{align}
where
\begin{align}
 q_1(x) &\equiv 16.0x^{-2}+0.00869x^{-1}-1.37+0.0842x-0.0477x^2+0.0340x^3, \nonumber\\
 q_2(x) &\equiv 32.0x^{-2}+0.0302x^{-1}-0.120+0.208x-0.229x^2+0.0322x^3, \nonumber\\
 q_3(x) &\equiv 128x^{-2}+0.395x^{-1}-0.946+0.836x-0.221x^2-0.00938x^3, 
\end{align}
and 
\begin{align}
 \tilde{Q}_1(x;l_{\rm max}) &= \sum_{l=1}^{l_{\rm max}}x^2l(l+1)(2l+1)K_2\left(x\sqrt{l(l+1)}\right), \nonumber\\
 \tilde{Q}_2(x;l_{\rm max}) &= \sum_{l=1}^{l_{\rm max}}x^3l^{3/2}(l+1)^{3/2}(2l+1)K_1\left(x\sqrt{l(l+1)}\right), \nonumber\\
 \tilde{Q}_3(x;l_{\rm max}) &= \sum_{l=1}^{l_{\rm max}}x^4l^2(l+1)^2(2l+1)K_0\left(x\sqrt{l(l+1)}\right). 
\end{align}

\begin{figure}[t]
  \begin{center}
    \includegraphics[scale=0.55]{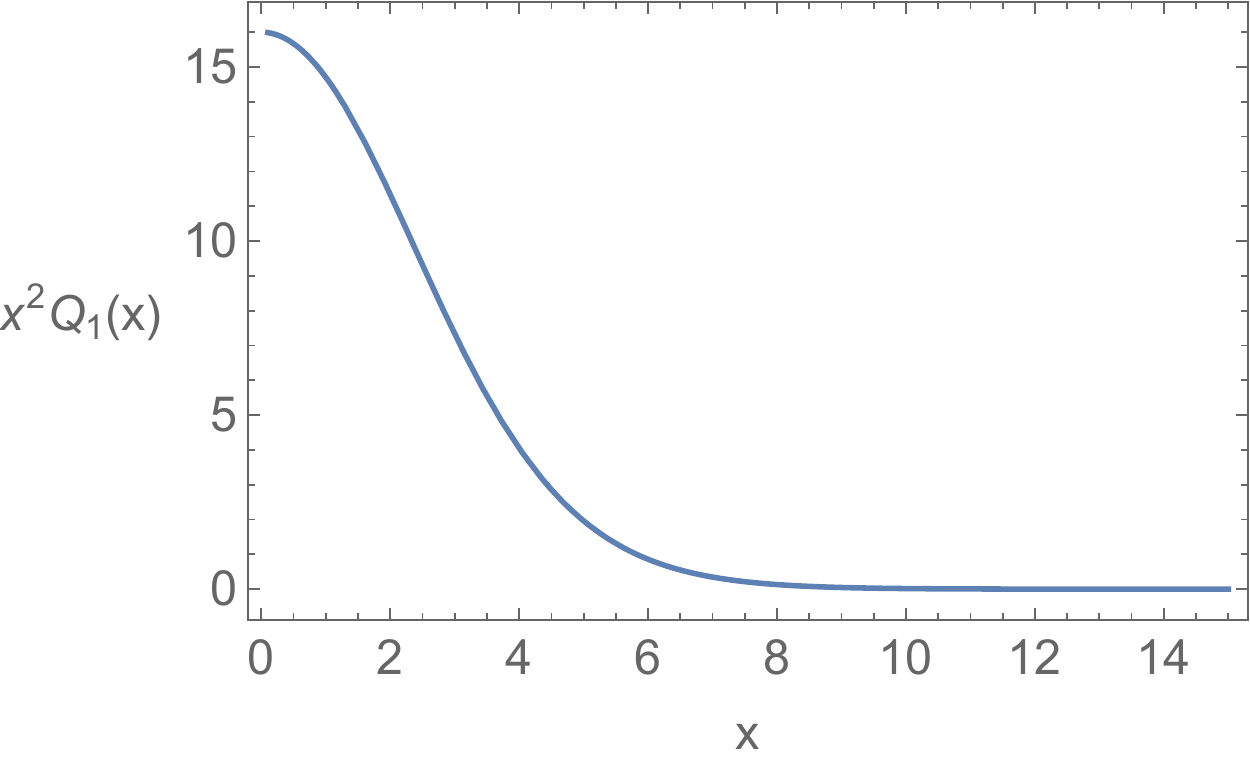} \hspace{10mm}
    \includegraphics[scale=0.55]{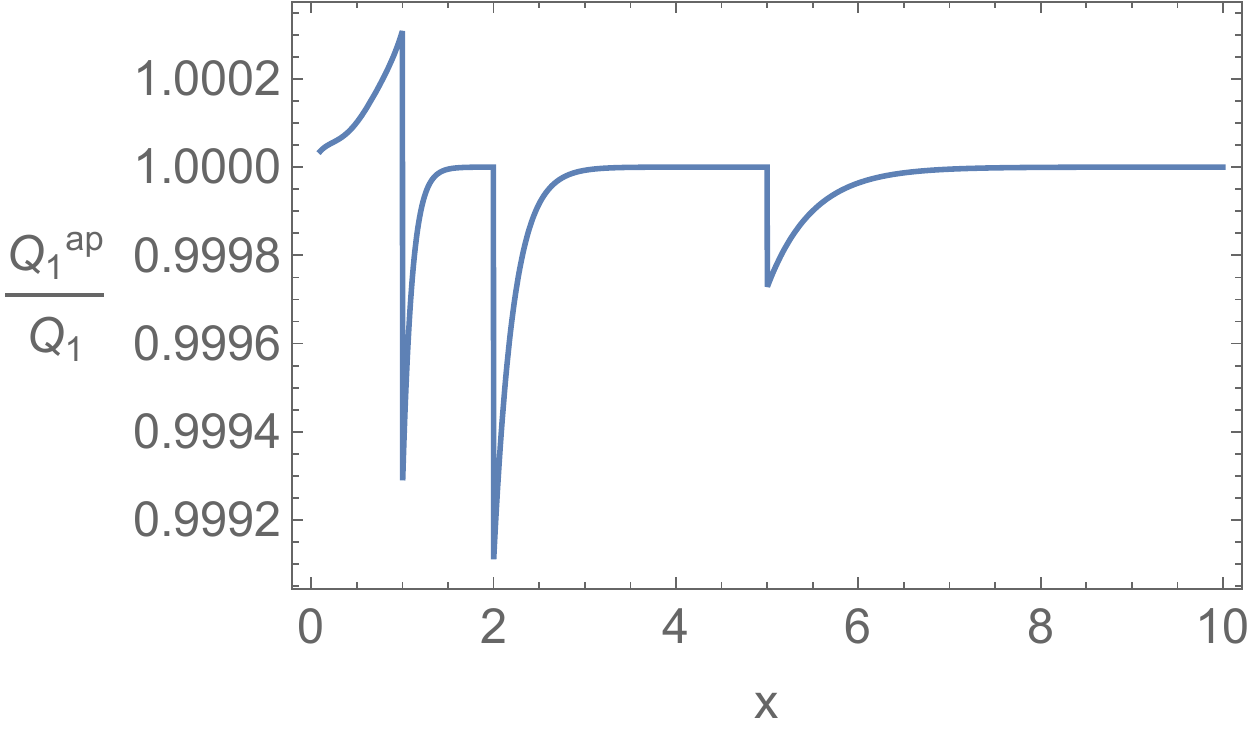} \\
    \includegraphics[scale=0.55]{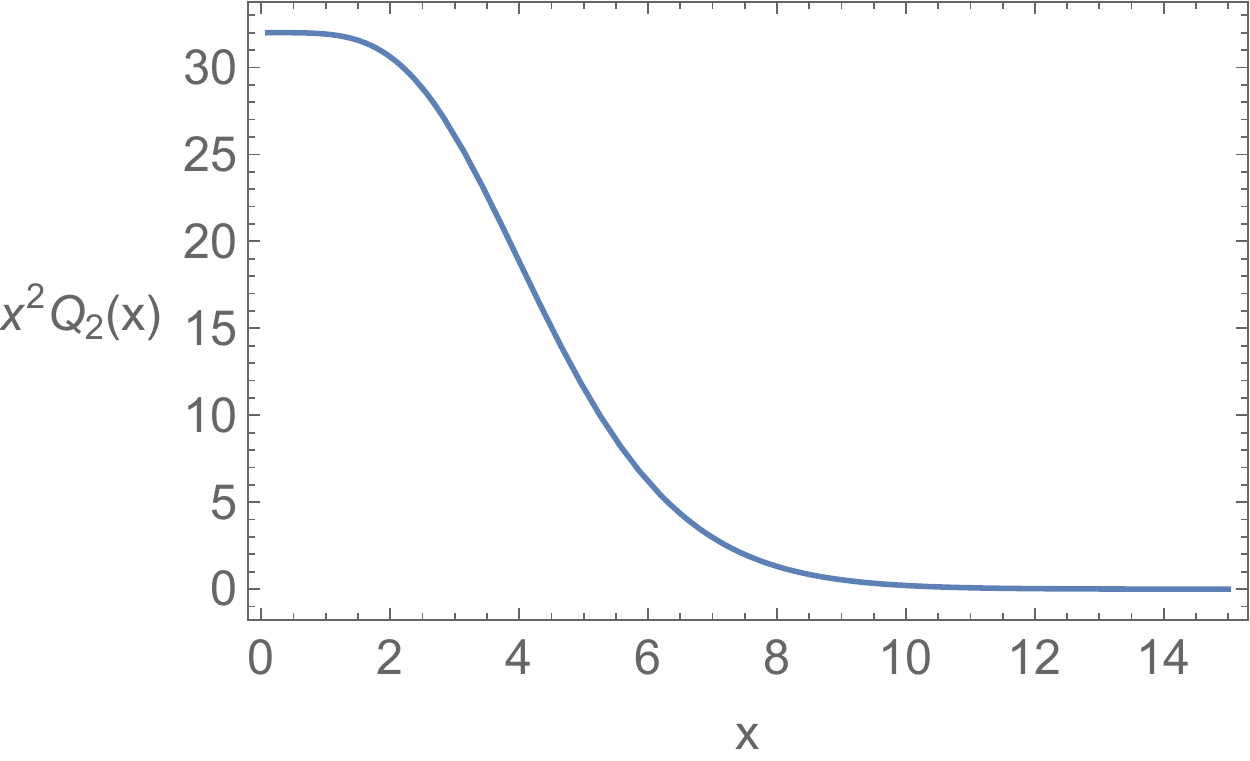} \hspace{10mm}
    \includegraphics[scale=0.55]{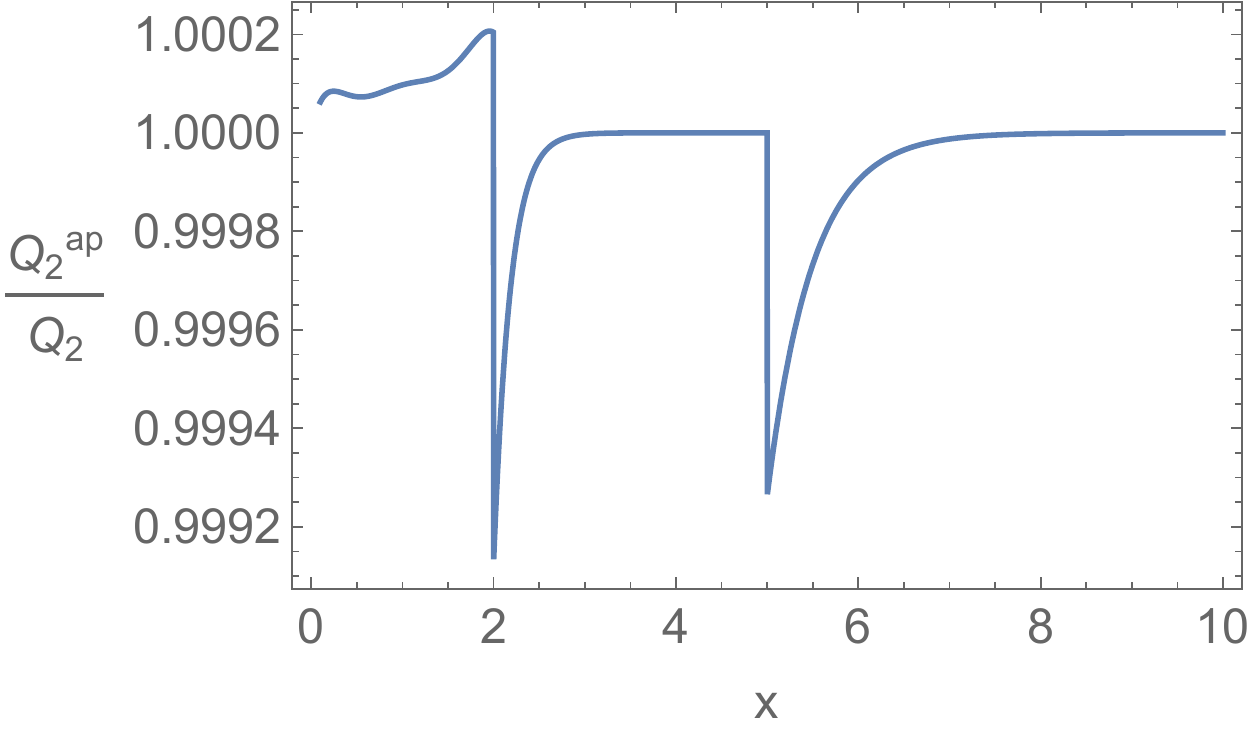} \\
    \includegraphics[scale=0.55]{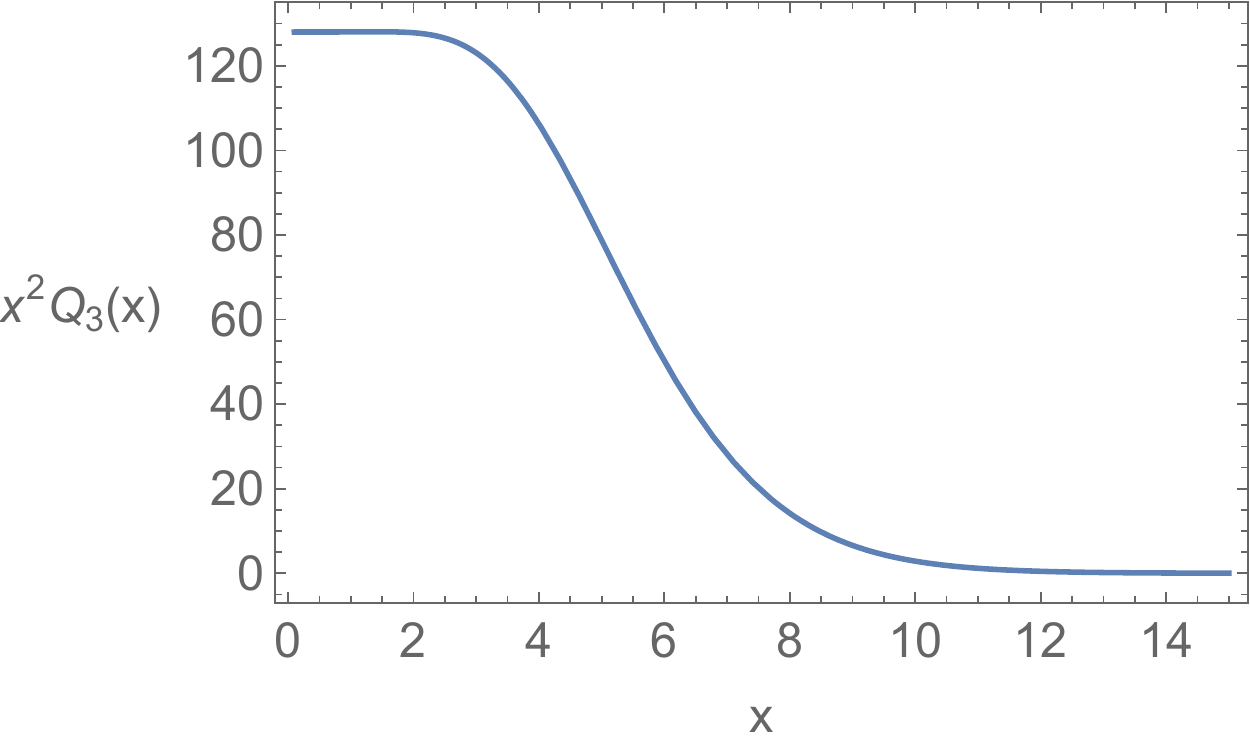} \hspace{10mm}
    \includegraphics[scale=0.55]{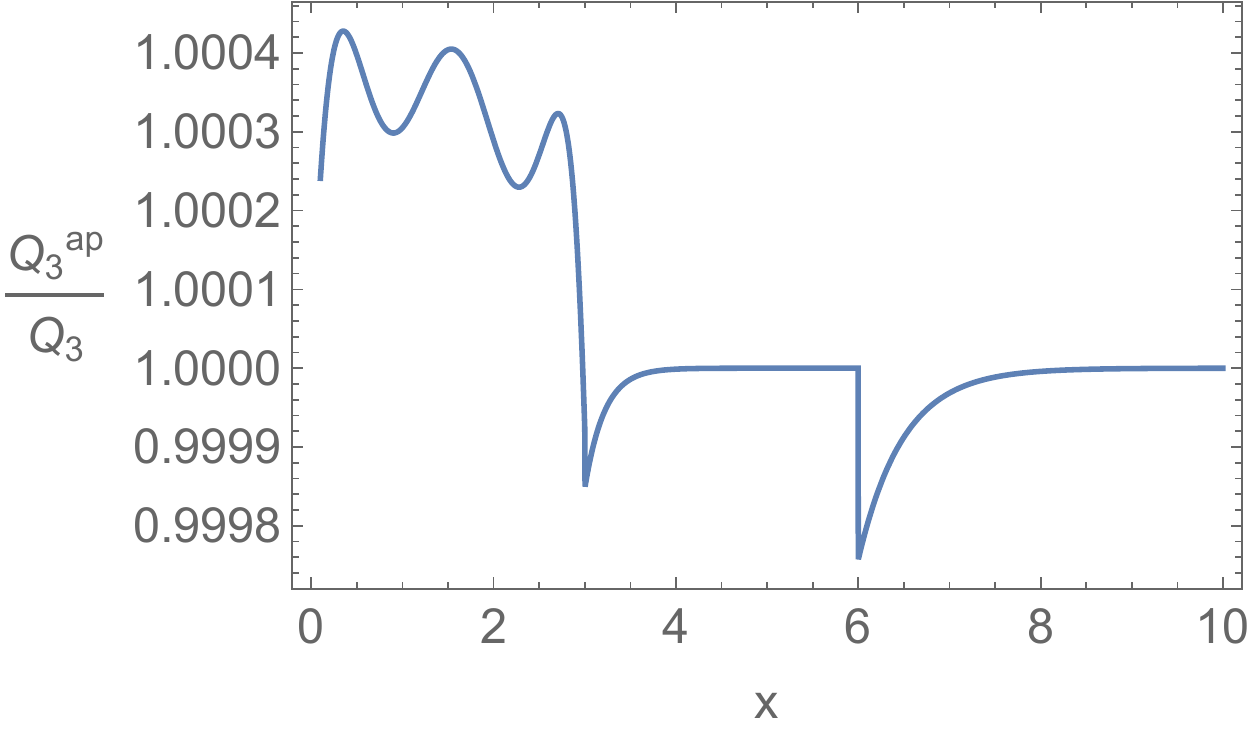}
  \end{center}
\caption{The profiles of $x^2Q_i(x)$ ($i=1,2,3$) (left plots), and the ratios to their approximate functions defined in (\ref{apQs}) (right plots). 
We have used $\tilde{Q}_i(x;300)$ for $Q_i(x)$ ($i=1,2,3$). }
    \label{profile:Qs}
\end{figure}

Fig.~\ref{profile:Qs} shows the profiles of $x^2Q_i(x)$ ($i=1,2,3$) and the ratios~$Q_i^{\rm ap}(x)/Q_i(x)$. 
The plots shows the above functions well approximate $Q_i(x)$ with the errors less than $10^{-3}$. 
We will use these approximate functions for the numerical calculations. 

\subsection{Effective potential}
\label{sec:3_2}

Before going to analyze the time evolution of the 6D thermal background, 
let us discuss the stabilization of the compact space~$S^2$. 
As we mentioned at the end of Sec.~\ref{subsec:SS}, the dilaton potential~(\ref{eq:SSscalar}) does not stabilize the radius of $S^2$. 
To fix it, 
we add the stabilization potential~$V_{\rm stab}(\sigma)$ to (\ref{eq:SSscalar}).  
Once the dilaton~$\sigma$ is stabilized to a finite value, the radion~$b$ is also fixed through the relation~(\ref{eq:SSvev}).

We expect that the stabilization potential~$V_{\rm stab}(\sigma)$ would be generated by dilaton-dependent quantum corrections allowed 
by the gauged $U(1)_R$ symmetry. 
It is quite important to derive $V_{\rm stab}(\sigma)$ from the string theory, 
but still a challenging issue. 
We leave the derivation of 6D gauged supergravity with the stabilization 
potential for a future work.\footnote{For earlier studies of the realization of lower-dimensional gauged supergravities from the string theory, 
see e.g., Ref. \cite{Cvetic:1999un}. 
The stabilization of moduli fields was also discussed in the context of 
4D low-energy effective field theory\cite{Aghababaie:2002be}.}

For the purpose of discussing the time evolution of the 5D space around and after the moduli stabilization, 
we do not need to know the complete functional form of $V_{\rm stab}(\sigma)$. 
The only relevant properties of $V_{\rm stab}(\sigma)$ are the position of the minimum and the curvature around it. 
Thus, we simply introduce the following form of $V_{\rm stab}(\sigma)$ 
as an approximation around the potential minimum. 
\begin{align}
V_{\rm stab}(\sigma) = \frac{m^2}{2} \left( \sigma-\sigma_\ast \right)^2, \label{V_stab}
\end{align}
where $m$ and $\sigma_\ast$ are real parameters.

The KK mass scale (or the compactification scale) is characterized by 
\begin{align}
 m_{\rm KK} &\equiv \frac{1}{b}. 
\end{align}
From (\ref{E_pl}), the first excited KK mass is $\sqrt{2}m_{\rm KK}$. 
Thus, there are two possible cases: 
\begin{itemize}
\item $m \gtrsim m_{\rm KK} $ \;\;\; (The moduli are already fixed in the 4D effective theory.) 
\item $m \ll m_{\rm KK}$ \;\;\; (The moduli stabilization can be treated in the 4D effective theory.) 
\label{mllmKK}
\end{itemize}
In the latter case, the dynamics of the moduli stabilization can be described within the 4D effective theory. 
From the 6D Lagrangian (\ref{eq:6DSeff}) and the potential~(\ref{V_stab}), the potential for the moduli is described by
\begin{align}
    V_{\rm mod}(b,\sigma) &= -\frac{1}{2}g^{mn}R_{mn}
 +\frac{g^2e^\sigma}{4}F^{mn}F_{mn} + 2e^{-\sigma} +V_{\rm stab}(\sigma) \nonumber\\
    &= -b^{-2}+\frac{e^{\sigma}}{8b^4}+2e^{-\sigma} +\frac{m^2}{2} \left( \sigma -\sigma_\ast\right)^2 \nonumber\\
 &= 2e^{-\sigma}\left(1-\frac{e^\sigma}{4b^2}\right)^2+\frac{m^2}{2}(\sigma-\sigma_*)^2. 
\label{eq:Veff}
\end{align}
Minimizing this potential, we find that both $b$ and $\sigma$ are stabilized at 
\begin{align}
    b_{\rm vac} &= \frac{e^{\sigma_\ast/2}}{2},
    \;\;\;\;\;\;
    \sigma_{\rm vac} = \sigma_\ast. 
\label{eq:vacuum}
\end{align}
Note that the potential energy at the minimum~(\ref{eq:vacuum}) is zero. 
Thus the geometry approaches the Minkowski spacetime at late times.

Recall that one direction in the moduli space~$\{b,\sigma\}$, which is specified by (\ref{eq:SSvev}), 
is already stabilized in the original Salam-Sezgin model. 
Now the other direction is also fixed by the stabilization potential $V_{\rm stab}(\sigma)$ whose typical scale is characterized by $m$. 
We assume that $V_{\rm stab}(\sigma)$ is generated at some scale well below the 6D Planck mass scale by some mechanism,
i.e., $m\ll 1$. 
Since we are interested in the behavior around the vacuum values~(\ref{eq:vacuum}), the size modulus~$b$ takes 
values of ${\cal O}(b_{\rm vac})$. 
Hence, the relation~$m\ll m_{\rm KK}$ indicates that $m^2b_{\rm vac}^2=m^2e^{\sigma_\ast}/4 \ll 1$.  
When $m\gtrsim m_{\rm KK}$, the moduli stabilization process should be analyzed in the context of the full 6D theory. 

\begin{figure}[H]
  \begin{center}
    \includegraphics[scale=0.5]{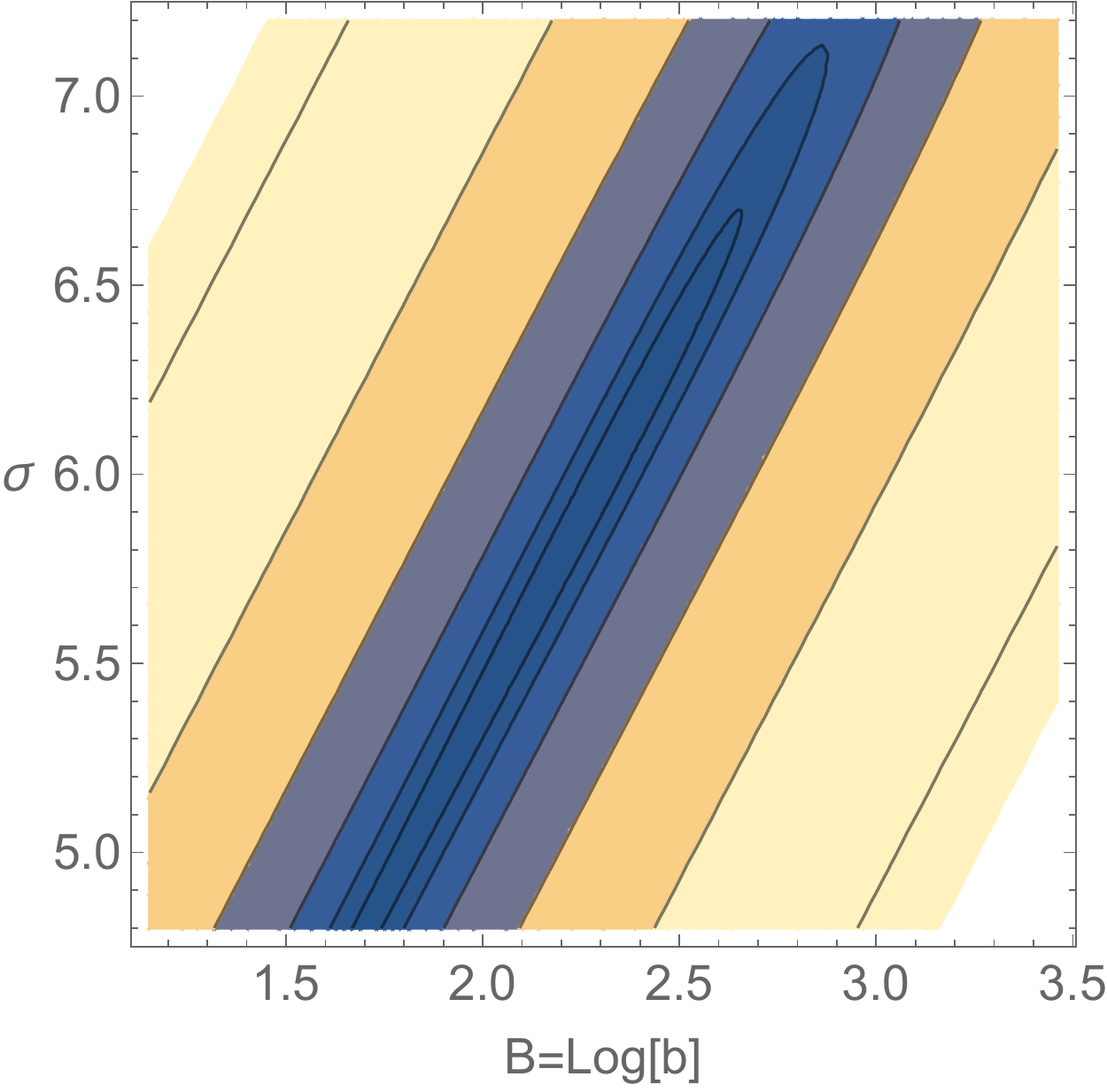} \hspace{5mm}
    \includegraphics[scale=0.48]{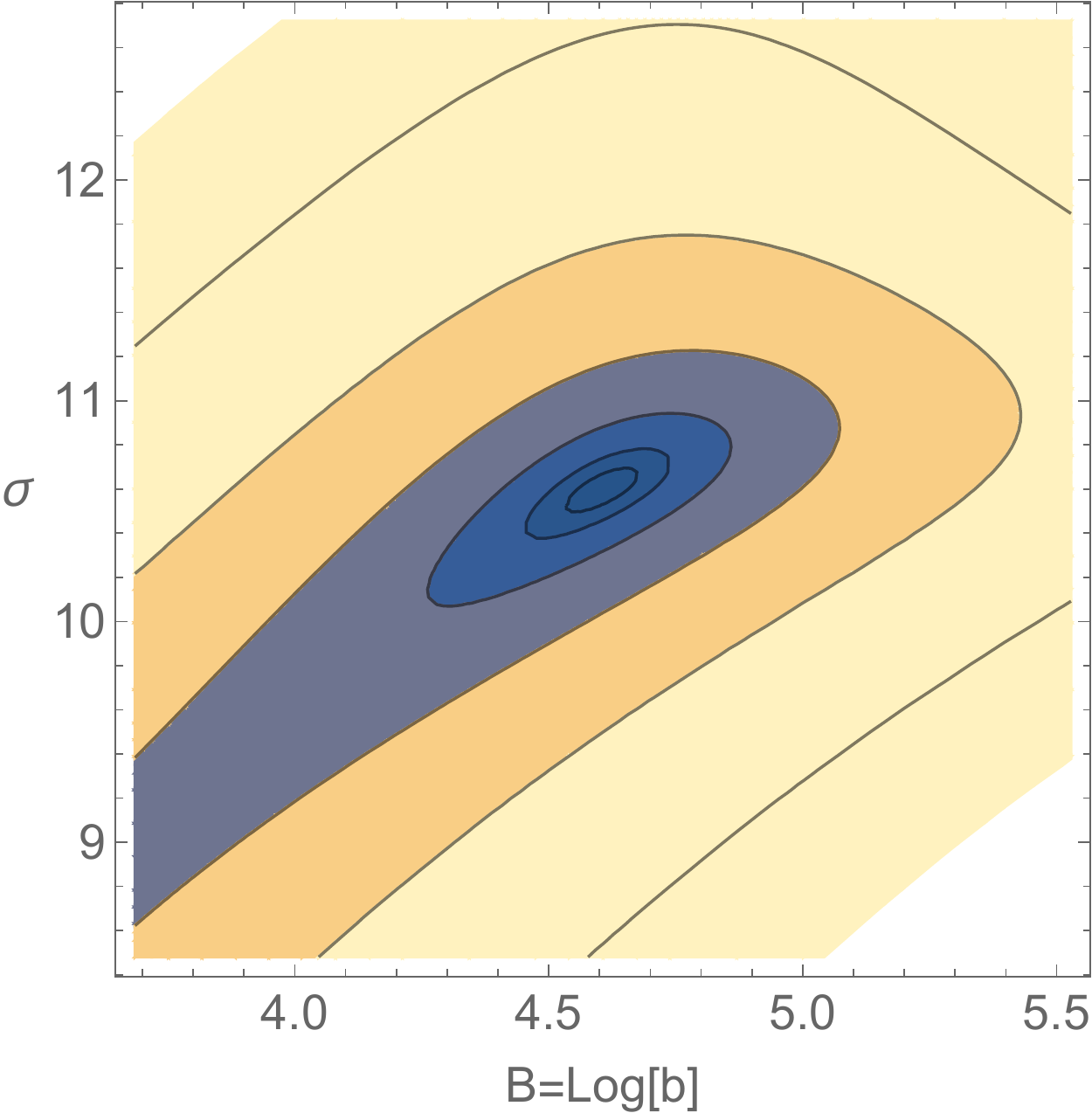}  \hspace{5mm}
    \includegraphics[scale=0.5]{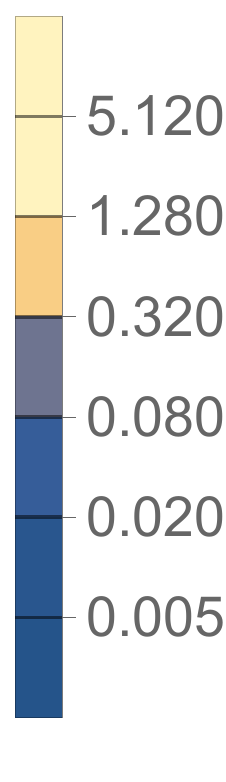}
  \end{center}
\caption{The moduli potential~$V_{\rm mod}(b,\sigma)$ with $(m,\sigma_\ast)=(10^{-2},6.0)$ (left plot) 
and $(m,\sigma_\ast)=(10^{-2},10.6)$ (right plot). 
For the left plot, the minimum is $(B,\sigma)=(2.3,6.0)$, and $m^2b_{\rm vac}^2=0.01$. 
For the right plot, they are $(4.6,10.6)$ and 1.0, respectively. }
    \label{fig:potential}
\end{figure}

\begin{figure}[H]
  \begin{center}
    \includegraphics[scale=0.5]{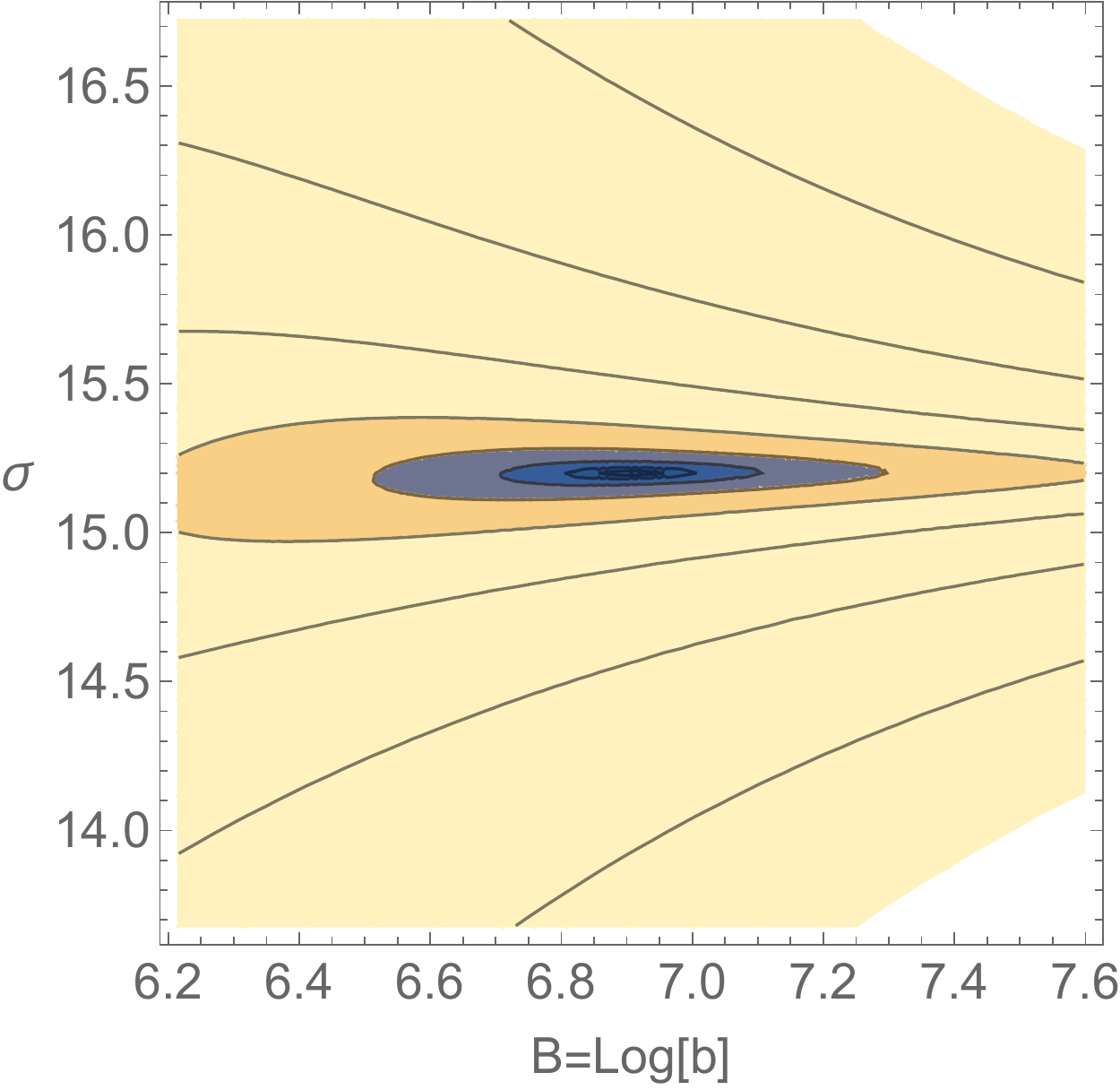} \hspace{5mm}
    \includegraphics[scale=0.5]{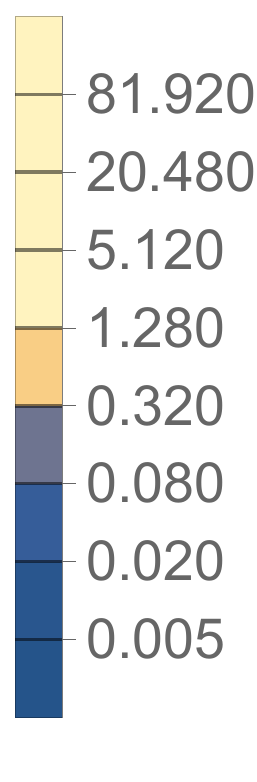}
  \end{center}
\caption{The moduli potential~$V_{\rm mod}(b,\sigma)$ with $(m,\sigma_\ast)=(10^{-2},15.2)$. 
The minimum is $(B,\sigma)=(6.8,15.2)$, and $b_{\rm vac}=1000$. }
    \label{fig:potential2}
\end{figure}

Fig. \ref{fig:potential} shows the moduli potential~$V_{\rm mod}(b,\sigma)$ for different values of $m^2b_{\rm vac}$. 
In the left plot, the stabilized values of the moduli are $(b_{\rm vac},\sigma_{\rm vac})=(10.0,6.0)$, and 
$m^2b_{\rm vac}^2=0.01\ll 1$. 
In this case, the two directions corresponding to the mass eigenvalues have hierarchically different curvatures. 
In the right plot, the minimum is $(b_{\rm vac},\sigma_{\rm vac})=(100,10.6)$, and $m^2b_{\rm vac}^2=1.0$. 
So the 4D effective theory analysis cannot be used in this case. 
Around the minimum, the moduli potential~$V_{\rm mod}$ has the form of
\begin{align}
 V_{\rm mod} &= -\frac{1}{2}(\delta b,\delta\sigma)
 \begin{pmatrix} 64e^{-2\sigma_\ast} & -16e^{-\frac{3}{2}\sigma_\ast} \\ -16e^{-\frac{3}{2}\sigma_\ast} & m^2+4e^{-\sigma_\ast} 
 \end{pmatrix}\begin{pmatrix} \delta b \\ \delta\sigma \end{pmatrix}+\cdots, 
 \label{bottom_V_mod}
\end{align}
where $\delta b \equiv b-b_{\rm vac}$ and $\delta\sigma \equiv \sigma-\sigma_{\rm vac}$, 
and the ellipsis denotes higher order terms in $\delta b$ or $\delta\sigma$. 
When $b_{\rm vac}$ is large enough, the mixing between $\delta b$ and $\delta\sigma$ is negligible 
as shown in Fig.~\ref{fig:potential2}. 
In this case, the curvature at the potential minimum along the $\delta\sigma$ direction is determined by the parameter~$m$, 
while that along the $\delta b$ direction is suppressed by $b_{\rm vac}^{-2}$.  

\section{Numerical analyses}
\label{sec:3_3}

\subsection{Evolution equations}
We are now ready for the numerical analysis of the background time evolution. 
Including the radiation contribution to the energy-momentum tensor~$T_{MN}$, 
which is expressed in terms of the thermodynamic quantities in Sec. \ref{sec:3_1}, 
the equations of motion~(\ref{EoM}) with the background ansatz~(\ref{NM:BGansatz})-(\ref{NM:BGansatz:3}) are summarized as follows:
\begin{align}
 &\frac{3\dot{a}^2}{a^2}+\frac{\dot{b}^2}{b^2}+\frac{6\dot{a}\dot{b}}{ab}+\frac{1}{b^2}-\frac{1}{2}\dot{\sigma}^2
 -\frac{e^\sigma}{8b^4}-2e^{-\sigma}-V_{\rm stab}(\sigma)-\rho_{\rm rad} = 0, \nonumber\\
 &\frac{2\ddot{a}}{a}+\frac{\dot{a}^2}{a^2}+\frac{2\ddot{b}}{b}+\frac{\dot{b}^2}{b^2}+\frac{4\dot{a}\dot{b}}{ab}
 +\frac{1}{b^2}+\frac{1}{2}\dot{\sigma}^2-\frac{e^\sigma}{8b^4}-2e^{-\sigma}-V_{\rm stab}(\sigma)+p_{\rm rad,3} = 0, \nonumber\\
 &\frac{3\ddot{a}}{a}+\frac{3\dot{a}^2}{a^2}+\frac{\ddot{b}}{b}+\frac{3\dot{a}\dot{b}}{ab}+\frac{1}{2}\dot{\sigma}^2
 +\frac{e^\sigma}{8b^4}-2e^{-\sigma}-V_{\rm stab}(\sigma)+p_{\rm rad,2} = 0, \nonumber\\
 &\ddot{\sigma}+\left(\frac{3\dot{a}}{a}+\frac{2\dot{b}}{b}\right)\dot{\sigma}+\frac{e^\sigma}{8b^4}
 -2e^{-\sigma}+\partial_\sigma V_{\rm stab}(\sigma) = 0, 
 \label{EOMs:abs}
\end{align}
where $\rho_{\rm rad}$, $p_{\rm rad,3}$ and $p_{\rm rad,2}$ are given 
by (\ref{rho_rad:exact}), (\ref{P_3:exact}) and (\ref{P_2:exact}) with (\ref{def:p23}), respectively. 
The first equation is the $(t,t)$-component of the Einstein equation. 
Since this does not contain the second order $t$-derivatives, it is regarded as a constraint on the initial conditions 
of the time-evolution equations.
The second and the third equations come from the diagonal components for the 3D non-compact space and the $S^2$ space, 
respectively. 
The other components of the Einstein equation vanish. 
The last equation is the dilaton equation of motion. 

For the radiation energy density~$\rho_{\rm rad}$ and the pressures~$p_{\rm rad,3}$ and $p_{\rm rad,2}$, 
we use the approximated forms in (\ref{expr:rho_rad}), and assume that the chemical potential~$\mu$ is negligible. 
Then, (\ref{EOMs:abs}) and (\ref{expr:dotbeta}) are rewritten as~\footnote{
The last equation in (\ref{eq:EOMsim}) can also be  obtained from the entropy-conservation condition~(\ref{conserve:cS_rad}) 
by neglecting the dependence of the chemical potential~$\mu$. 
} 
\begin{align}
 \begin{split}
 \ddot{A} =& -\frac{1}{4}\left(9\dot{A}^2-\dot{B}^2+2\dot{A}\dot{B}-e^{-2B}+\frac{1}{2}\dot{\sigma}^2
 +\frac{3}{8}e^{\sigma-4B}-2e^{-\sigma}-V_{\rm stab}(\sigma)-p_{\rm rad,3}^{\rm ap}+2p_{\rm rad,2}^{\rm ap}\right),
 \end{split} \nonumber\\
 \ddot{B} =& -\frac{1}{4}\left(-3\dot{A}^2+7\dot{B}^2+6\dot{A}\dot{B}+3e^{-2B}+\frac{1}{2}\dot{\sigma}^2 
 -\frac{5}{8}e^{\sigma-4B}-2e^{-\sigma}-V_{\rm stab}(\sigma)+3p_{\rm rad,3}^{\rm ap}-2p_{\rm rad,2}^{\rm ap}\right),
 \nonumber\\
 \ddot{\sigma} =& -\left(3\dot{A}+2\dot{B}\right)\dot{\sigma}-\frac{1}{8}e^{\sigma-4B}+2e^{-\sigma}-\partial_\sigma V_{\rm stab}(\sigma), 
 \nonumber\\
 \frac{\dot{\beta}}{\beta} =& 
 \frac{3\dot{A}\left\{\pm 8{\rm Li}_4(\pm 1)+4Q_1+Q_2\right\}
 +\dot{B}\left(2Q_2+Q_3\right)}{\pm 24{\rm Li}_4(\pm 1)+12Q_1+5Q_2+Q_3}, 
 \label{eq:EOMsim}
\end{align}
where the arguments of $Q_i$ ($i=1,2,3$) are $\beta e^{-B}$, and 
\begin{align}
 A &\equiv \ln a, \;\;\;\;\;
 B \equiv \ln b, 
\end{align}
and
\begin{align}
 \rho_{\rm rad}^{\rm ap} &= \frac{g_{\rm dof}e^{-2B}}{8\pi^3\beta^4}
 \left\{\pm 6{\rm Li}_4(\pm 1)+3Q_1+Q_2\right\}, \nonumber\\
 p_{\rm rad,3}^{\rm ap} &= \frac{g_{\rm dof}e^{-2B}}{8\pi^3\beta^4}
 \left\{\pm 2{\rm Li}_4(\pm 1)+Q_1\right\}, \;\;\;\;\;
 p_{\rm rad,2}^{\rm ap} = \frac{g_{\rm dof}e^{-2B}}{16\pi^3\beta^4}Q_2, 
\end{align}
with the constraint: 
\begin{align}
 &3\dot{A}^2+\dot{B}^2+6\dot{A}\dot{B}+e^{-2B}-\frac{1}{2}\dot{\sigma}^2-\frac{1}{8}e^{\sigma-4B}
 -2e^{-\sigma}-V_{\rm stab}(\sigma)-\rho_{\rm rad}^{\rm ap} = 0. 
 \label{cstrt}
\end{align}
These are the evolution equations we will numerically solve in the following.

Notice that the above evolution equations do not depend on the non-derivative value of $A$. 
Namely, if $A_{\rm sol}(t)$ is a solution, $A_{\rm sol}(t)+A_0$ ($A_0$: constant) is also a solution. 
This is because only the relative ratio of the scale factors~$a(t)$ at different times is physically meaningful.\footnote{ 
The value of $a$ at a specific time can be freely changed by the 3D coordinate~$x^i$ ($i=1,2,3$).}
In contrast, the value of $b(t)$ has the physical meaning, i.e., the radius of the extra space~$S^2$.\footnote{ 
Recall that we have chosen the coordinate~$y$ so that $r=1$. }
In fact, the evolution equations explicitly depend on the non-derivative $B(t)$.

\subsection{Parameter choice and initial values}
In the following numerical calculations, we choose the stabilization scale~$m$ and the degrees of freedom 
for the radiation particles~$g_{\rm dof}$ as
\begin{align}
 m &= 0.01\,\,\mbox{(except for Fig.\,\ref{bsgm:m001})}, \;\;\;\;\;
 g_{\rm dof} = 100. 
 \label{choice:ini_values}
\end{align}
The parameter~$\sigma_\ast$ is used to control the stabilized moduli values in (\ref{eq:vacuum}). 

The initial values at $t=0$ are chosen as
\begin{align}
 \beta(0) &= 10\,\,\mbox{(except for Fig.\,\ref{plots:winv1-bt100})}, \nonumber\\
 a(0) &= 1, \;\;\;\;\;
 b(0) = b_{\rm vac}-\delta, \;\;\;\;\;
 \sigma(0) = \sigma_{\rm vac}+\delta, \;\;\;\;\;
 \nonumber\\
 \dot{a}(0) &= \frac{1}{\sqrt{3}}\sqrt{V_{\rm mod}(b,\sigma)+\rho_{\rm rad}^{\rm ap}}, \;\;\;\;\;
 \dot{b}(0) = \dot{\sigma}(0) = 0, 
 \label{initial_values}
\end{align}
where $\{b_{\rm vac},\sigma_{\rm vac}\}$ are the stabilized moduli values defined in (\ref{eq:vacuum}), 
$V_{\rm mod}(b,\sigma)$ is defined in (\ref{eq:Veff}), and $\delta$ is chosen as 0.1. 
The value of $\dot{a}(0)$ is determined by the constraint~(\ref{cstrt}).

\subsection{Numerical results}

\subsubsection{Moduli stabilization}
\begin{figure}[H]
  \begin{center}
    \includegraphics[scale=0.7]{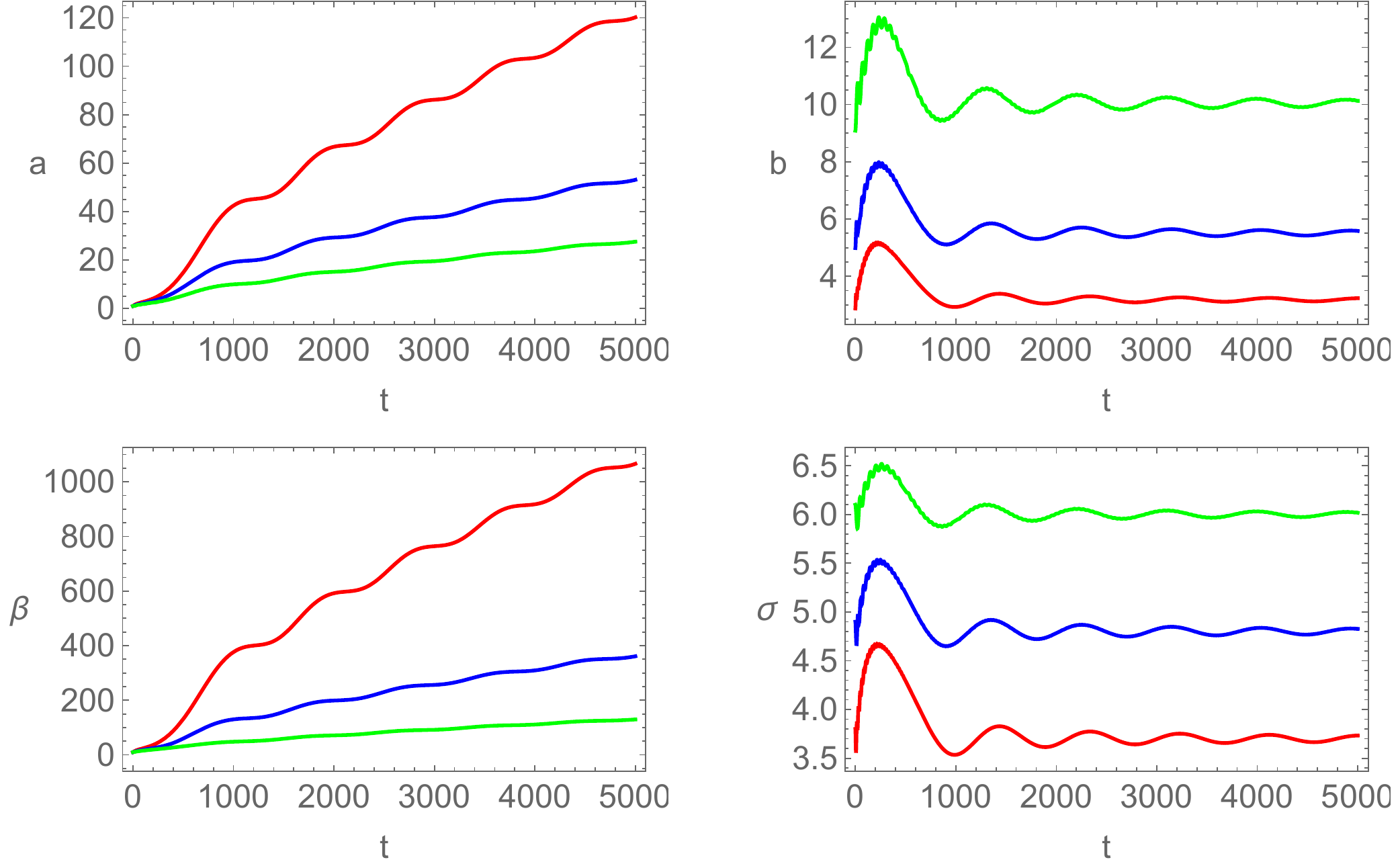} 
  \end{center}
\caption{Time evolution of the scale factors~$\{a,b\}$, the inverse temperature~$\beta$, and the dilaton~$\sigma$. 
The value of $\sigma_\ast$ is chosen from the bottom to the top as 
$\sigma_\ast=3.7$, 4.8, 6.0, which correspond to $m^2b_{\rm vac}^2=0.001$, 0.003, 0.01, respectively. 
The other parameters are chosen as (\ref{initial_values}).}
\label{case_small}
\end{figure}

\begin{figure}[H]
  \begin{center}
    \includegraphics[scale=0.7]{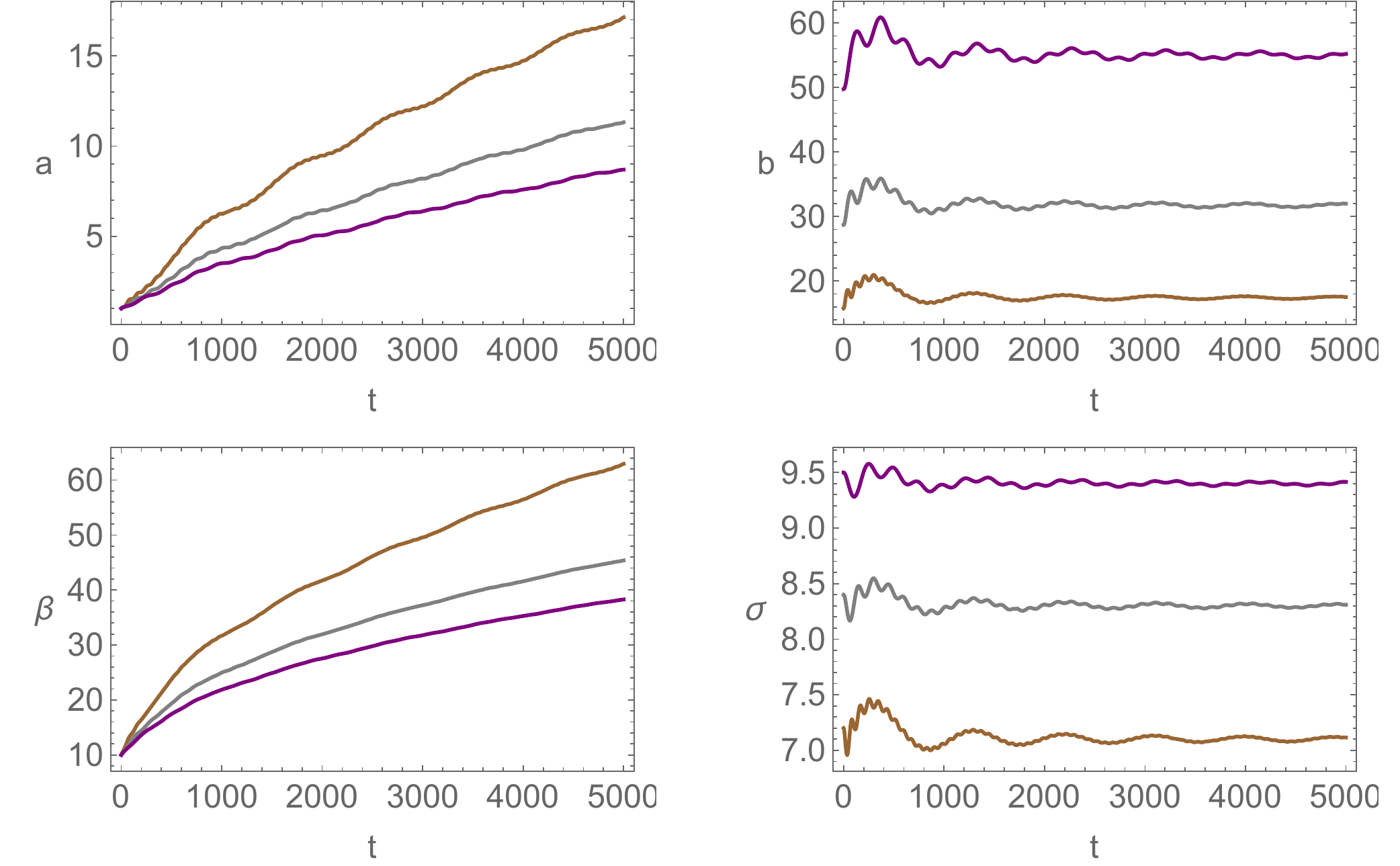}
  \end{center}
\caption{Time evolution of the scale factors~$\{a,b\}$, the inverse temperature~$\beta$, and the dilaton~$\sigma$. 
The value of $\sigma_\ast$ is chosen from the bottom to the top as 
$\sigma_\ast=7.1$, 8.3, 9.4, which correspond to $m^2b_{\rm vac}^2=0.03$, 0.1, 0.3, respectively.  
The other parameters are chosen as (\ref{initial_values}).}
\label{case_large}
\end{figure}

Figs.~\ref{case_small} and \ref{case_large} show the time evolution of the scale factors~$\{a,b\}$, the inverse temperature~$\beta$, 
and the background dilaton value~$\sigma$ for various values of $m^2b_{\rm vac}^2$.  
From these plots, we can see that the moduli~$b$ and $\sigma$ settle down at the stabilized values until $t\sim 1000$ 
in all cases. 
Namely, the time scale required for the moduli stabilization does not depend on the value of $\sigma_\ast$ (or $b_{\rm vac}$). 
We can also see that the moduli oscillate around the stabilized values in (\ref{eq:vacuum}). 
This consists of two oscillations with different periodicities, corresponding to the two eigenvalues of $\partial_\varphi\partial_\chi V_{\rm mod}$ 
($\varphi,\chi=b,\sigma$) at the minimum of the potential. 
As shown in Fig.~\ref{fig:potential}, these periodicities are hierarchical when $m^2b_{\rm vac}^2\ll 1$ 
while they are comparable when $m^2b_{\rm vac}^2={\cal O}(1)$. 

\begin{figure}[H]
  \begin{center}
    \includegraphics[scale=0.7]{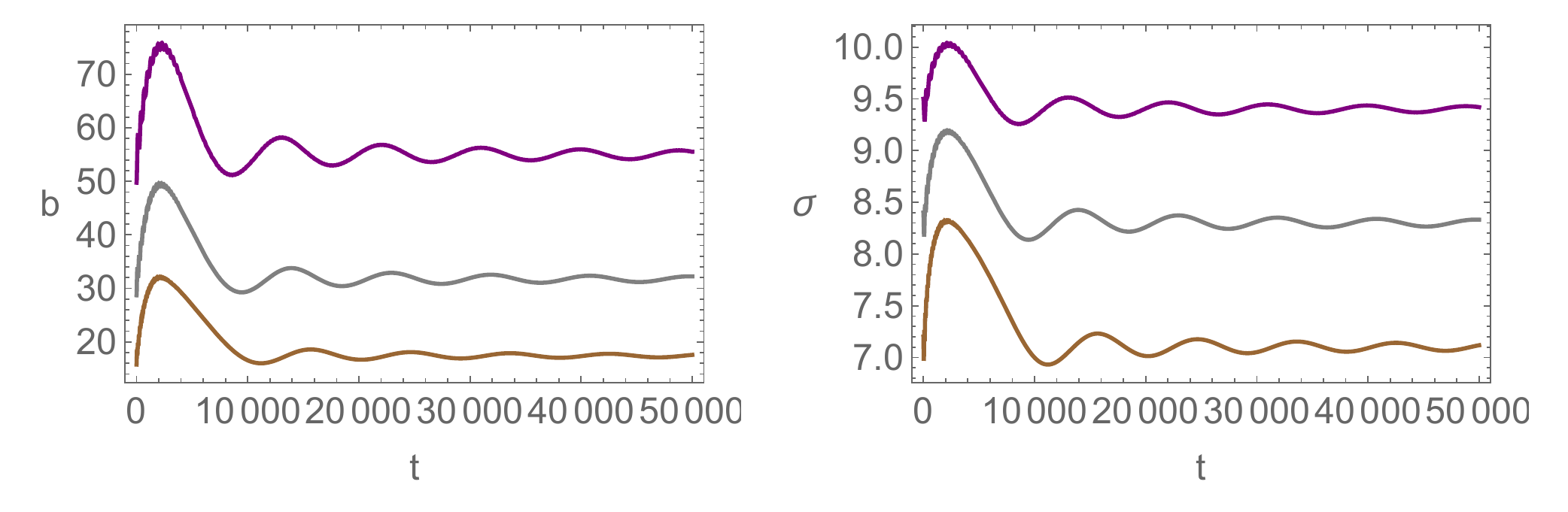}
  \end{center}
\caption{Time evolution of the scale factor~$b$ and the dilaton~$\sigma$ in the case of $m=0.001$. 
The value of $\sigma_\ast$ is chosen from the bottom to the top as 
$\sigma_\ast=7.1$, 8.3, 9.4, which correspond to $m^2b_{\rm vac}^2=3\times 10^{-4}$, $1\times 10^{-3}$, $3\times 10^{-3}$, respectively. 
The other parameters are chosen as (\ref{initial_values}). }
\label{bsgm:m001}
\end{figure}

Fig.~\ref{bsgm:m001} shows the time evolution of the moduli in the case of $m=0.001$. 
We can see that the time scale of the moduli stabilization is 10 times larger than the case of $m=0.01$ 
in Figs.~\ref{case_small} and \ref{case_large}. 
This indicates that this time scale~$t_{\rm stab}$ is estimated as about $10m^{-1}$. 

We should note that the above properties are not sensitive to the choice of the initial values. 
In fact, similar results are obtained even if we choose $\delta=0$ in (\ref{initial_values}) 
since $\dot{a}(0)$ is non-zero due to the existence of the radiation.

\subsubsection{Energy densities of radiation and moduli oscillation}
At late times when $b$ is stabilized, its time derivative can be neglected. 
Then, the Einstein equations: 
\begin{align}
 R_{tt}-\frac{1}{2}g_{tt}R &= T_{tt}, \nonumber\\
 R_{ij}-\frac{1}{2}g_{ij}R &= T_{ij}, 
 \label{Einstein_eq}
\end{align}
which correspond to the first and the second equations in (\ref{EOMs:abs}), 
become 
\begin{align}
 \frac{3\dot{a}^2}{a^2}+\frac{1}{b^2} 
 &= \frac{1}{2}\dot{\sigma}^2+\frac{e^\sigma}{8b^4}+2e^{-\sigma}+V_{\rm stab}(\sigma)+\rho_{\rm rad} = \rho_{\rm tot}, \nonumber\\
 \frac{2\ddot{a}}{a}+\frac{\dot{a}^2}{a^2}+\frac{1}{b^2} 
 &= -\frac{1}{2}\dot{\sigma}^2+\frac{e^\sigma}{8b^4}+2e^{-\sigma}+V_{\rm stab}(\sigma)-p_{\rm rad,3} = -p_{\rm tot,3}. 
 \label{latetime:Eeq}
\end{align}
The `constant' terms~$1/b^2$ in the LHSs come from the scalar curvature (in the second terms in LHSs of (\ref{Einstein_eq})). 
For the purpose of seeing the correspondence to the 4D Einstein equations, 
it is convenient to redefine the energy density and the 3D pressure as
\begin{align}
 \tilde{\rho}_{\rm tot} &\equiv \rho_{\rm tot}-\frac{1}{b^2} = \frac{1}{2}\dot{\sigma}^2+V_{\rm mod}(b,\sigma)+\rho_{\rm rad}, \nonumber\\
 \tilde{p}_{\rm tot,3} &\equiv p_{\rm tot,3}+\frac{1}{b^2} = \frac{1}{2}\dot{\sigma}^2-V_{\rm mod}(b,\sigma)+p_{\rm rad,3},  
\end{align}
where $V_{\rm mod}(b,\sigma)$ is defined in (\ref{eq:Veff}). 
Then (\ref{latetime:Eeq}) has the same form as the 4D Friedmann equations. 
In the following, we use these quantities even when $b$ has not been stabilized yet, i.e., it is time-dependent. 

\begin{figure}[H]
  \begin{center}
    \includegraphics[scale=0.6]{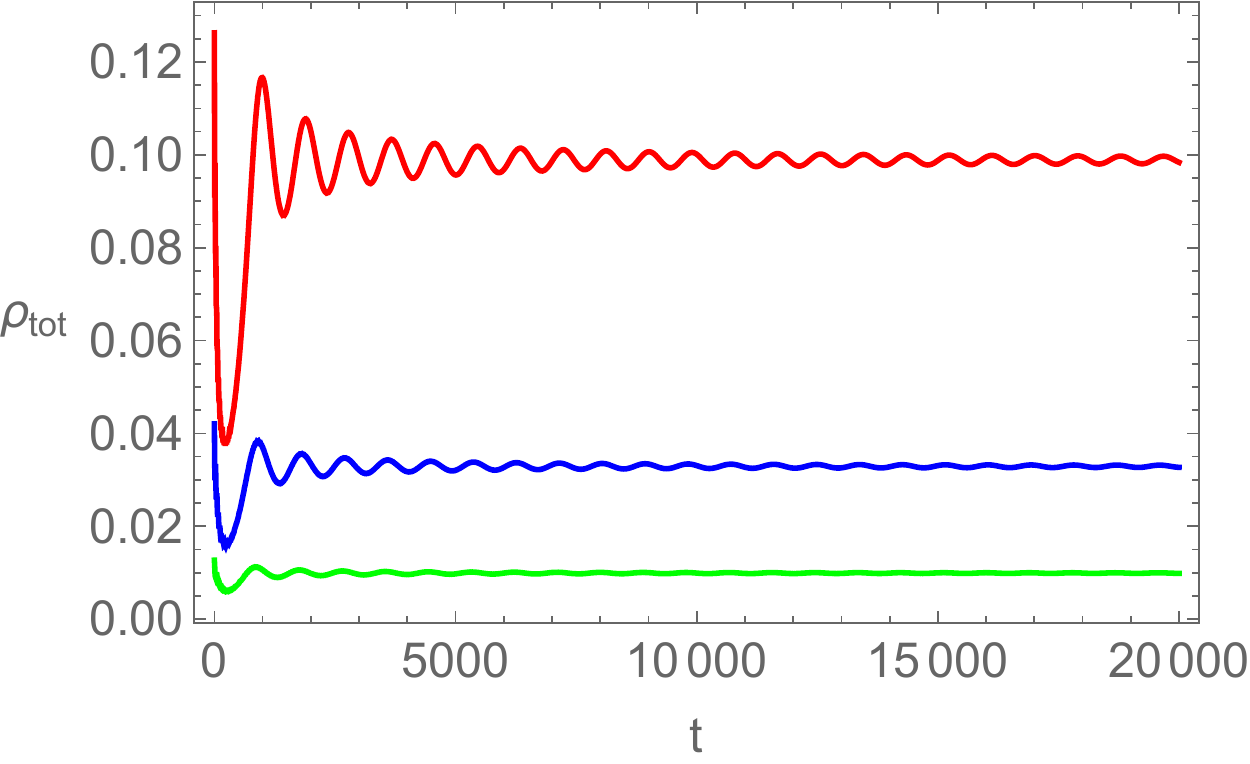} \;\;
    \includegraphics[scale=0.6]{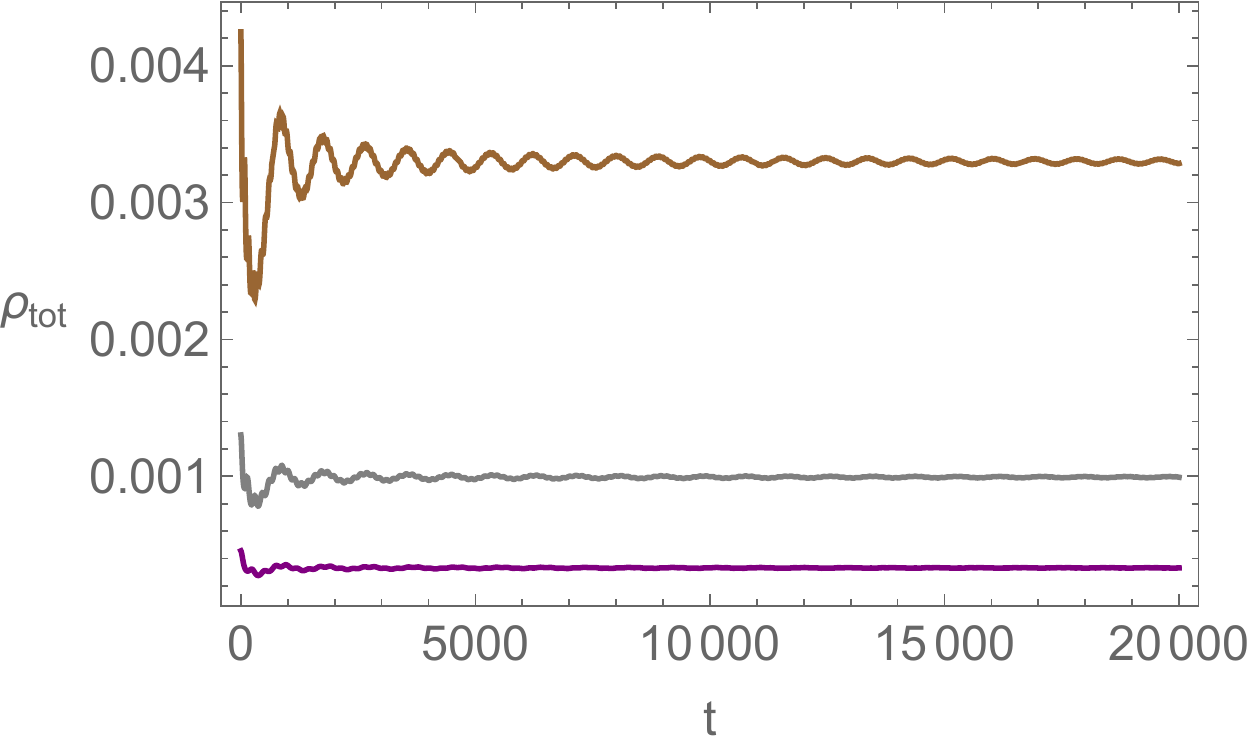}
  \end{center}
\caption{Time evolution of $\rho_{\rm tot}$ defined in (\ref{Tcomp}). 
The parameter is chosen from the bottom to the top as $\sigma_\ast=3.7$, 4.8, 6.0, which correspond to 
$m^2b_{\rm vac}^2=0.001$, 0.003, 0.01 in the left plot, and $\sigma_\ast=7.1$, 8.3, 9.4, which correspond to
$m^2b_{\rm vac}^2=0.03$, 0.1, 0.3 in the right plot, respectively. 
The other parameters are chosen as (\ref{initial_values}).}
\label{plot:rhotot}
\end{figure}

Fig.~\ref{plot:rhotot} shows the time evolution of $\rho_{\rm tot}$ defined in (\ref{Tcomp}). 
We can see that it converges to a constant value 
at the same time scale as the moduli stabilization (i.e., $\sim 10m^{-1}$). 
We have checked that the radiation contribution~$\rho_{\rm rad}$ is rapidly damped and 
becomes negligible at the moduli-stabilized time~$t_{\rm stab}$. 
The constant value which $\rho_{\rm tot}$ approach is read off as $1/b_{\rm vac}^2$. 
Hence $\tilde{\rho}_{\rm tot}$ is damped to zero at late times. 

\begin{figure}[H]
  \begin{center}
    \includegraphics[scale=0.64]{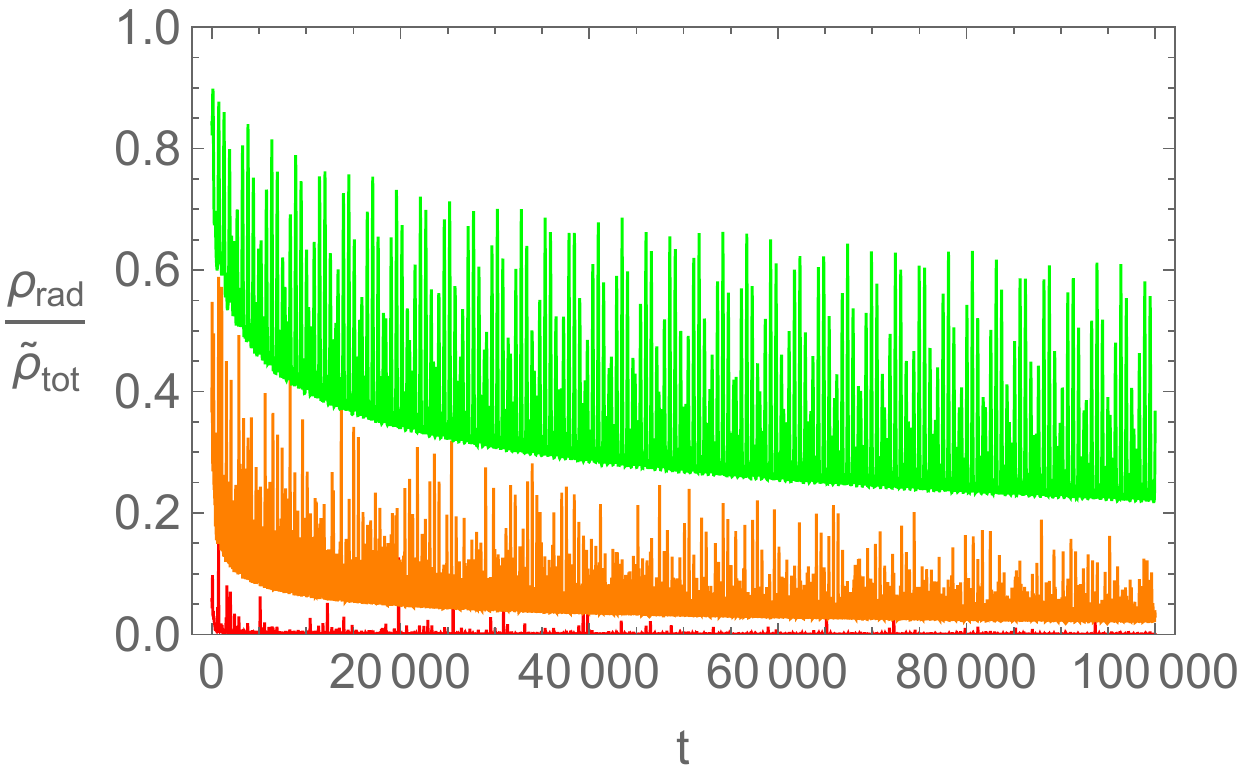} \;\;
    \includegraphics[scale=0.63]{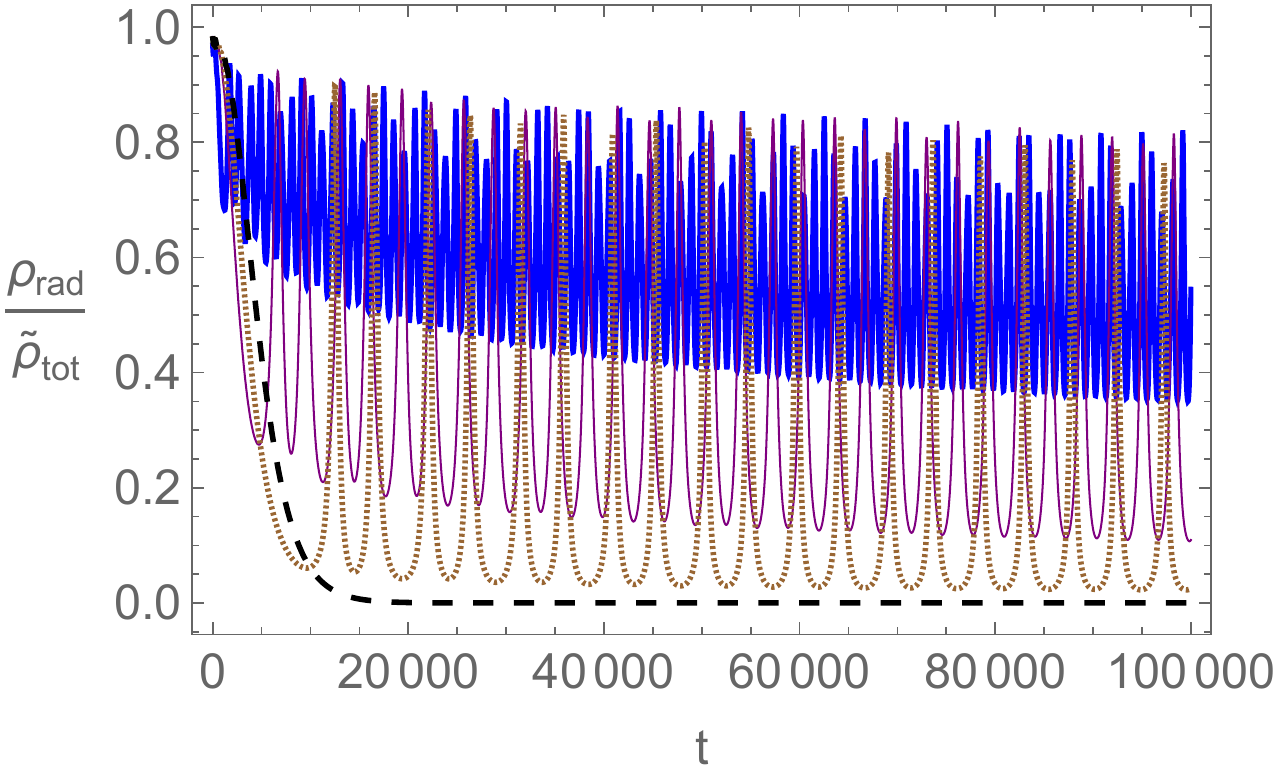}
  \end{center}
\caption{Time evolution of the ratio~$\rho_{\rm rad}/\tilde{\rho}_{\rm tot}$ for various values of $\sigma_*$. 
The parameter is chosen from the bottom to the top as $\sigma_\ast=6.0$ (red), 8.3 (orange), 10.6 (green), 
which correspond to $b_{\rm vac}=10$, 31.7, 100 in the left plot, 
and $\sigma_\ast=12.9$ (blue), 15.2 (purple thin), 16.0 (brown fine dashed), 16.8 (black coarse dashed), 
which correspond to $b_{\rm vac}=316,1000,1490, 2223$ in the right plot, respectively. 
The other parameters are chosen as (\ref{initial_values}). }
\label{plot:rtorho}
\end{figure}

Fig.~\ref{plot:rtorho} shows the time evolution of the ratio of the radiation energy density~$\rho_{\rm rad}$ 
to the total energy density~$\tilde{\rho}_{\rm tot}$. 
In the case that $m^2b_{\rm vac}^2\ll 1$, the moduli stabilization can be described in the 4D effective theory. 
In such a case, it is known that the moduli oscillation around the potential minimum behaves as a non-relativistic matter, 
and thus its contribution to the energy density dominates over that of the radiation 
because $\rho_{\rm mod}\propto a^{-3}$ and $\rho_{\rm rad}\propto a^{-4}$, 
where $\rho_{\rm mod}\equiv \tilde{\rho}_{\rm tot}-\rho_{\rm rad}$ is the contribution of the moduli 
to the energy density. 
In fact, we can see that $\rho_{\rm rad}/\tilde{\rho}_{\rm tot}$ is rapidly damped 
when $\sigma_\ast=6.0$, i.e., $m^2b_{\rm vac}^2=0.01$. 
However, in the case that $m^2b_{\rm vac}^2\gtrsim 1$, the situation changes. 
We cannot describe the moduli stabilization process from the viewpoint of the 4D effective theory any more. 
From the plots, we can see that the radiation contriubtion to the total energy density becomes non-negligible 
as the stabilized size of $S^2$ is increased (the left plot in Fig.~\ref{plot:rtorho}).  
For $b_{\rm vac}=1000$, for example, the radiation energy density oscillates but remains to be larger than 40\% of the total one 
until about $100t_{\rm stab}$. 

\begin{figure}[H]
  \begin{center}
    \includegraphics[scale=0.65]{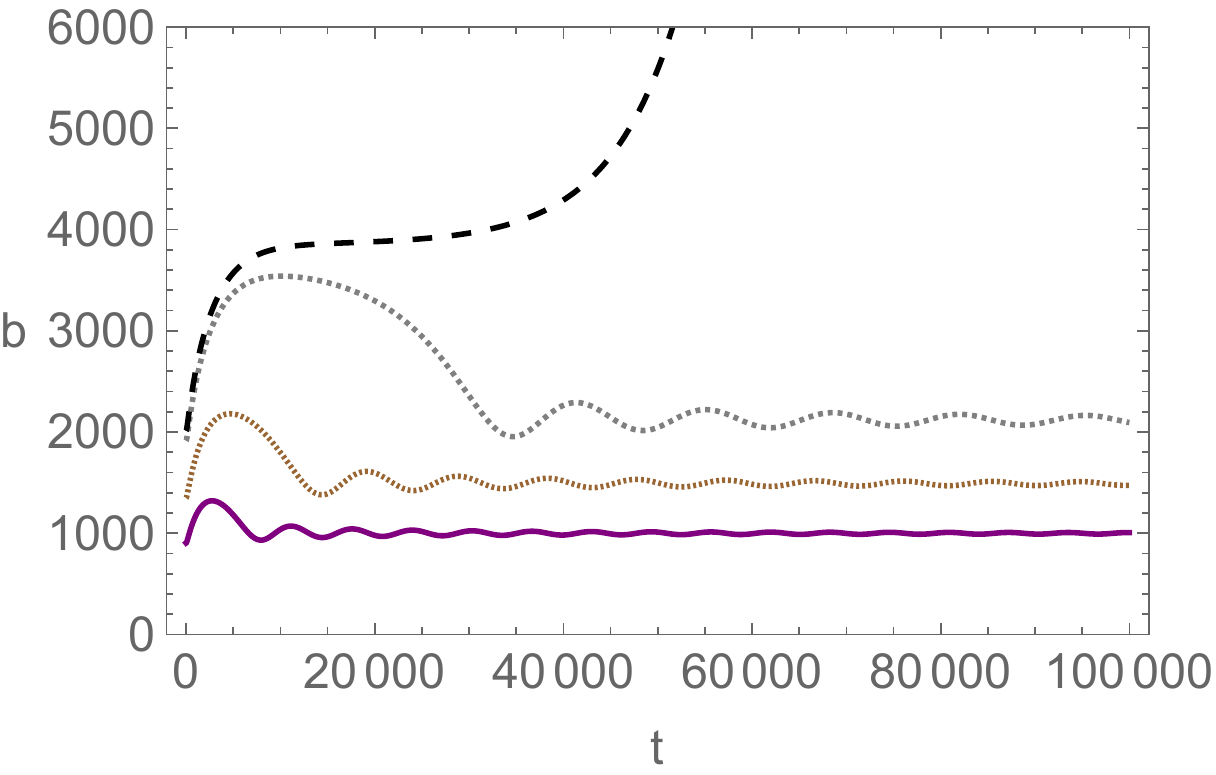} \;\;
    \includegraphics[scale=0.65]{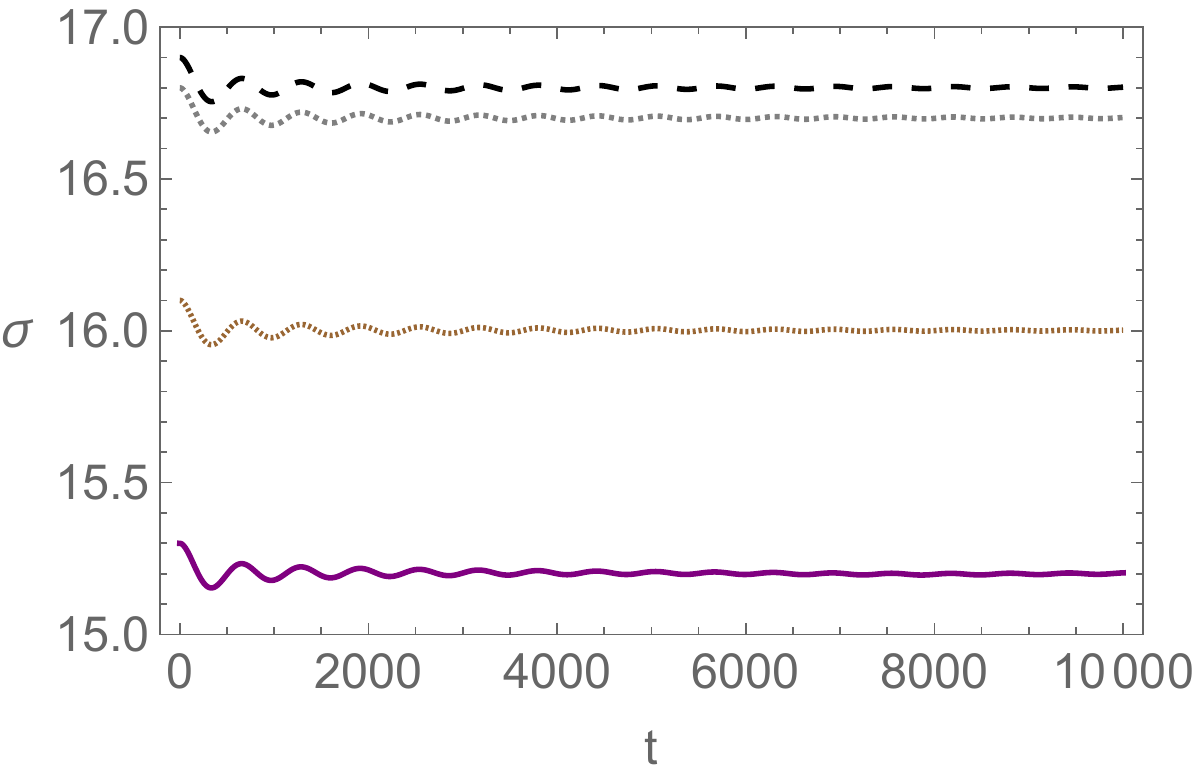}
  \end{center}
\caption{Time evolutions of $b$ (left plot) and $\sigma$ (right plot). 
The parameter~$\sigma_*$ is chosen from the bottom to the top as $\sigma_\ast=15.2$ (purple), 16.0 (brown fine dashed), 
16.7 (gray fine dashed), 16.8 (black coarse dashed), which correspond to $b_{\rm vac}=1000,1490, 2115, 2223$, respectively. 
The other parameters are chosen as (\ref{initial_values}). }
\label{plot:large_bsgm}
\end{figure}

For $b_{\rm vac}>1000$, the lower values of the oscillation for this ratio start to decrease, 
and $\rho_{\rm rad}$ becomes negligible at late times when $b_{\rm vac}\geq 2220$ (the right plot in Fig.~\ref{plot:rtorho}). 
This is because the modulus~$b$ is no longer stabilized 
and continues to increase for $b_{\rm vac}\geq 2220$ (see Fig.~\ref{plot:large_bsgm}). 
Namely, $\tilde{\rho}_{\rm tot}$ is not damped to zero, but approaches the value~$1/b_{\rm vac}^2$. 
Note that the potential curvature along the $b$-direction 
at the minimum~$\partial_b^2V_{\rm mod}(b_{\rm vac},\sigma_{\rm vac})$ 
becomes smaller as $b_{\rm vac}$ increases, as mentioned at the end of Sec.~\ref{sec:3}. 
Thus if $b_{\rm vac}$ is large enough, the modulus~$b$ cannot fall into the bottom of the potential~$V_{\rm mod}$ 
due to the Hubble friction.\footnote{
The dilaton~$\sigma$ falls into the potential minimum because $\partial_\sigma^2V_{\rm mod}(b_{\rm vac},\sigma_{\rm vac})$ 
is bounded from below by $m^2$. 
} 
In such a case, the non-vanishing vacuum energy density forces to expand the universe. 
As a result, the size modulus~$b$ increases further. 
We should note that this behavior depends on the structure of the moduli potential, and is not a general property 
of the moduli stabilization. 
If we assume a stabilization mechanism that also works for larger stabilized value~$b_{\rm vac}$, 
it is expected that the situation such that $\rho_{\rm rad}/\tilde{\rho}_{\rm tot}\simeq 1$ is realized. 

In the standard 4D cosmology, the 3D space expands as $a(t)\propto (t-t_0)^{2/3}$ for the matter-dominated universe 
while $a(t)\propto (t-t_0)^{1/2}$ for the radiation dominated one. 
Due to the non-negligible contribution of the radiation to the energy density, 
the 3D space expands slower than the matter-dominated case. 
In order to see a rough behavior of the scale factor~$a(t)$, we fit its numerical data with the function: 
\begin{align}
 a(t) &= N_a(t-t_0)^u, 
 \label{fit:a}
\end{align}
where $N_a$, $u$ and $t_0$ are real constants. 
The fitted values for various $b_{\rm vac}$ are listed in Table~\ref{fitting}. 
We can see that the power~$u$ is close to $2/3$
when the size of the compact space is small. 
However, its value decreases until $b_{\rm vac}=316$, 
and then turns to increase due to the destabilization of the size modulus~$b$ mentioned above. 
\begin{table}[h]
\centering
\begin{tabular}{|c||c|c|c|}
\hline
$b_{\rm vac}$ & $N_a$ & $u$ & $t_0$ \\ 
\hline\hline
10 & 0.09825 & 0.6633 & 210.8 \\
\hline
31.6 & 0.04073 & 0.6590 & -146.4 \\
\hline
100 & 0.03466 & 0.6243 & -325.1 \\
\hline
316 & 0.03817 & 0.6061 & -264.1 \\
\hline
500 & 0.03738 & 0.6122 & -181.8 \\
\hline
1000 & 0.03477 & 0.6405 & 270.7 \\
\hline
\end{tabular}
\caption{The fitted values of $N_a$, $u$ and $t_0$ in (\ref{fit:a}) for various values of $b_{\rm vac}$. }
\label{fitting}
\end{table}

\subsubsection{Equation of state for radiation}
\begin{figure}[H]
  \begin{center}
    \includegraphics[scale=0.6]{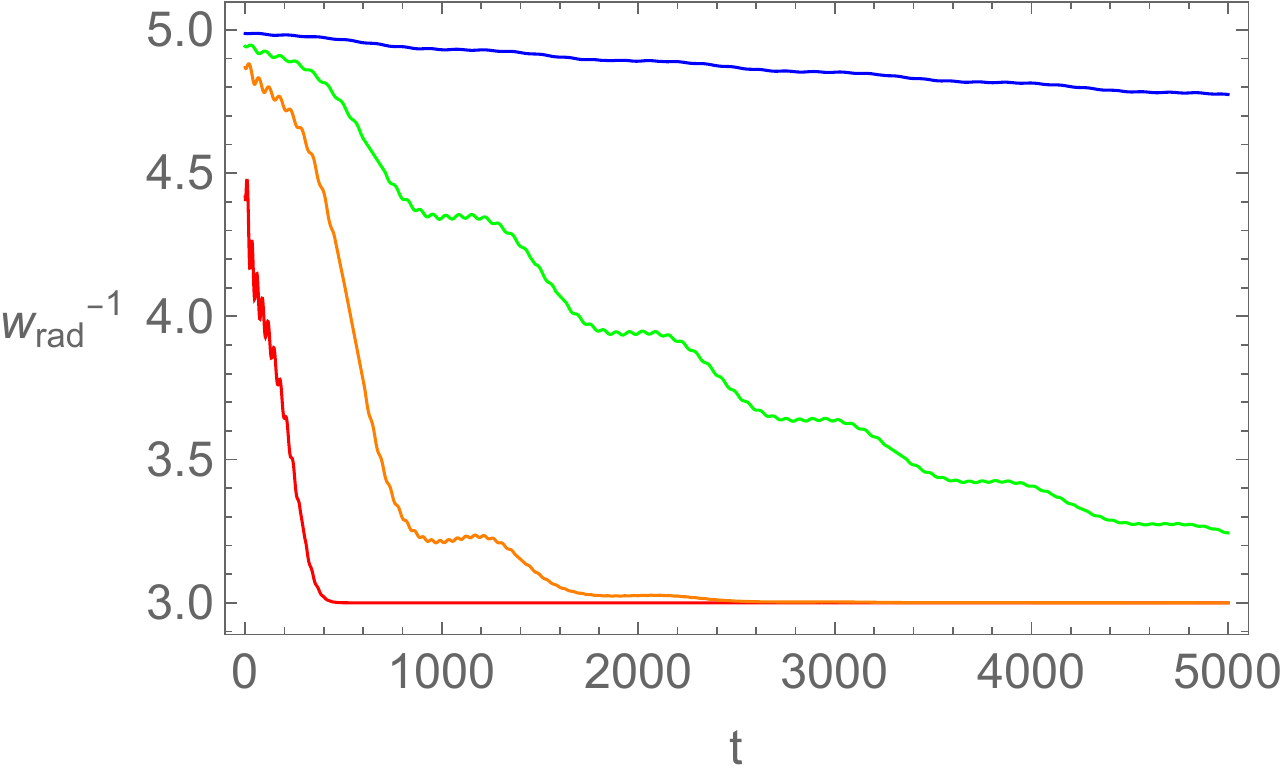} \;\;
    \includegraphics[scale=0.65]{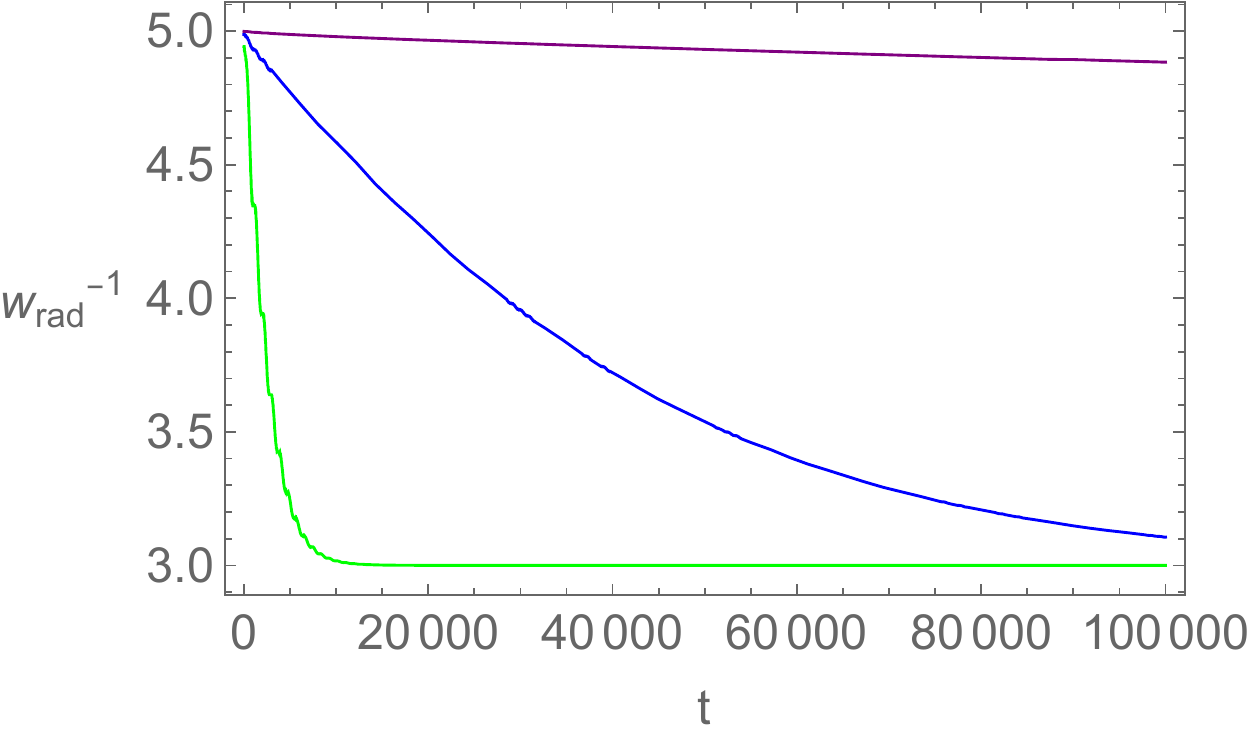}
  \end{center}
\caption{Time evolution of $w_{\rm rad}^{-1}$ for various values of $\sigma_*$. 
The parameter is chosen from the bottom to the top as $\sigma_\ast=4.6$ (red), 6.0 (orange), 6.8 (green), 8.3 (blue), 
which correspond to $b_{\rm vac}=5.0$, 10.0, 15.0, 31.7 in the left plot, and $\sigma_\ast=6.8$ (green), 8.3 (blue), 10.6 (purple), 
which correspond to $b_{\rm vac}=15.0$, 31.7, 100 in the right plot. 
The other parameters are chosen as (\ref{initial_values}). }
\label{plots:winv}
\end{figure}

Finally, we will see the equation of state for the radiation~$w_{\rm rad}$. 
As mentioned in Sec.~\ref{sec:3_1}, $w_{\rm rad}^{-1}$ measures the dimension of the space 
that the radiation feels. 
Fig.~\ref{plots:winv} shows the time evolution of $w_{\rm rad}^{-1}$ 
for various values of $\sigma_\ast$ (or $b_{\rm vac}$). 
From these plots, we can see that it takes a longer time until $w_{\rm rad}^{-1}$ settles down at the 3D space value~3 
as $b_{\rm vac}$ increases. 
Namely, when the size of $S^2$ is set to be large ($mb_{\rm vac}\gg 1$), the radiation feels the extra dimensions for a long time 
even after the moduli are stabilized. 
Also from this aspect, the 4D effective theory should not be used to analyze the stabilization procedure in such a case. 

\begin{figure}[H]
  \begin{center}
    \includegraphics[scale=0.65]{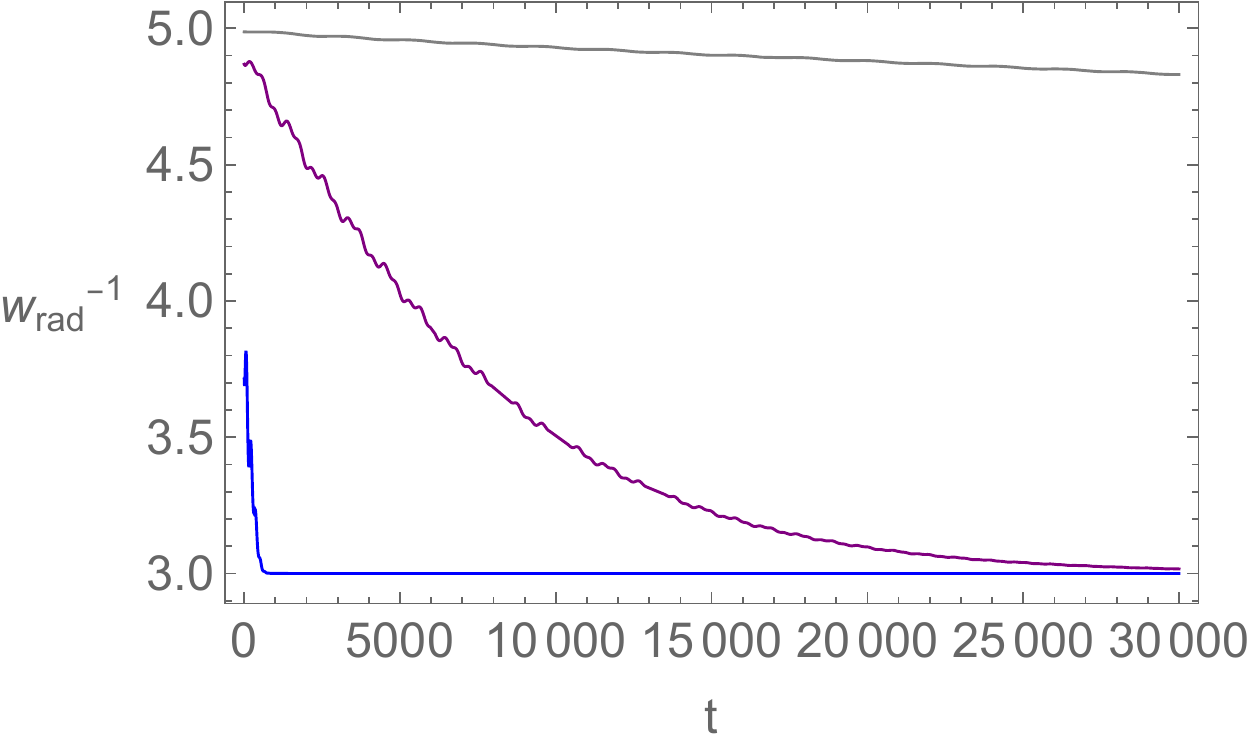} 
  \end{center}
\caption{Time evolution of $w_{\rm rad}^{-1}$ for various values of $\sigma_*$ in the case of $\beta(t=0)=100$. 
The parameter is chosen from the bottom to the top as $\sigma_\ast=8.3$ (blue), 10.6 (purple), 12.9 (gray) which correspond to 
$b_{\rm vac}=31.7$, 100, 316, respectively. 
The other parameters are chosen as (\ref{initial_values}). }
\label{plots:winv1-bt100}
\end{figure}

Since $w_{\rm rad}^{-1}$ is a monotonically decreasing function of $\beta/b$ (see Fig.~\ref{fig:winv}), 
it approaches 3 more rapidly if we choose a larger initial value of $\beta$. 
Fig.~\ref{plots:winv1-bt100} shows the time evolution of $w_{\rm rad}^{-1}$ in the case of $\beta(t=0)=100$. 

\subsection{4D Planck unit}
So far we have worked in the 6D Planck unit~$M_6=1$, where $M_6$ is the 6D Planck mass. 
Since the 4D Planck mass~$M_4$ is defined from the current gravitational coupling after the size of the compact space has been 
stabilized, the relation between $M_4$ and $M_6$ is given by 
\begin{align}
 M_4^2 &= 4\pi (b_{\rm vac}r)^2M_6^4 = 4\pi b_{\rm vac}^2M_6^2, 
\end{align}
where $b_{\rm vac}$ defined in (\ref{eq:vacuum}) is the stabilized value of $b$. 
Note that we have chosen the coordinate radius~$r$ as the 6D Planck length, i.e., $r=1/M_6$. 
Namely, the values of all dimensionful quantities in the previous subsections are measured in the unit of $M_6$. 
For example, since the inverse temperature~$\beta$, the stabilization mass scale~$m$ and the dilaton~$\sigma$ 
have mass-dimensions $-1$, 1 and 1, respectively, they should be understood as 
\begin{align}
 \beta &\to \frac{\beta}{M_6} = \beta\sqrt{\frac{4\pi b_{\rm vac}^2}{M_4^2}} = \frac{2\sqrt{\pi}b_{\rm vac}\beta}{M_4} 
 = \frac{\beta^{(4)}}{M_4}, \nonumber\\
 m &\to mM_6 = \frac{mM_4}{2\sqrt{\pi}b_{\rm vac}} = m^{(4)}M_4, \nonumber\\
 \sigma &\to \sigma M_6 = \frac{\sigma M_4}{2\sqrt{\pi}b_{\rm vac}} = \sigma^{(4)}M_4, 
\end{align}
where $\beta^{(4)}$, $m^{(4)}$ and $\sigma^{(4)}$ are the values measured in the unit of $M_4$. 
Namely, if we change the unit of the mass scale from $M_6$ to $M_4$, their values are changed as~\footnote{
The parameter~$\sigma_*$ denotes the value measured by $M_6$. 
}
\begin{align}
 \beta^{(4)} &= 2\sqrt{\pi}b_{\rm vac}\beta = \sqrt{\pi}e^{\sigma_*/2}\beta, \nonumber\\
 m^{(4)} &= \frac{m}{2\sqrt{\pi}b_{\rm vac}} = \frac{e^{-\sigma_*/2}}{\sqrt{\pi}}m, \nonumber\\
 \sigma^{(4)} &= \frac{\sigma}{2\sqrt{\pi}b_{\rm vac}} = \frac{e^{-\sigma_*/2}}{\sqrt{\pi}}\sigma. 
\end{align}
As for the dimensionless quantities such as the scale factors~$a$ and $b$, their values are unchanged. 
The physical radius of the compact space~$S^2$ is 
\begin{align}
 br = \frac{b}{M_6} = \frac{2\sqrt{\pi}b_{\rm vac}}{M_4}b. 
\end{align}
At late times after the moduli are stabilized, this approaches 
\begin{align}
 \frac{R_{\rm phys}^{(4)}}{M_4} \equiv \frac{2\sqrt{\pi}b_{\rm vac}^2}{M_4}, 
\end{align}
where $R_{\rm phys}$ is the value of the stabilized radius measured by $M_4$. 

\begin{figure}[H]
  \begin{center}
    \includegraphics[scale=0.6]{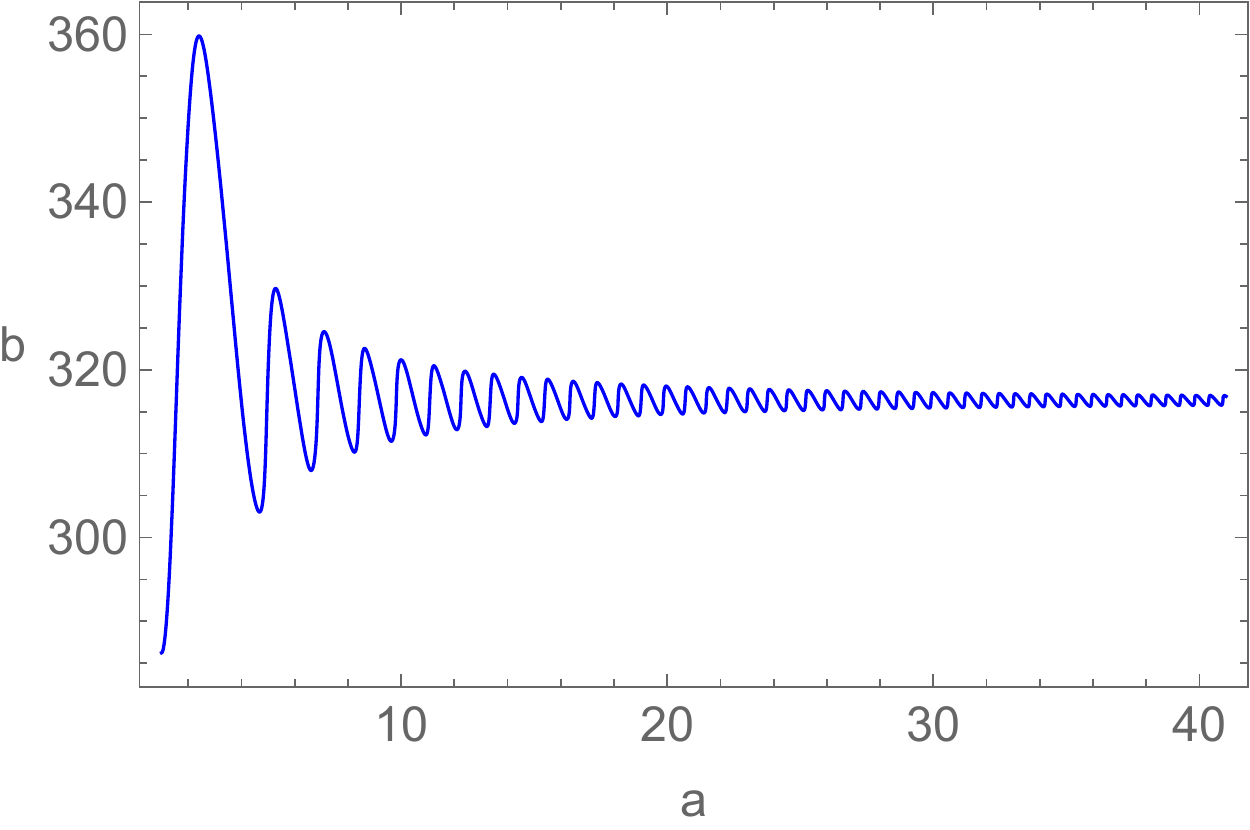} \;\;\;
    \includegraphics[scale=0.65]{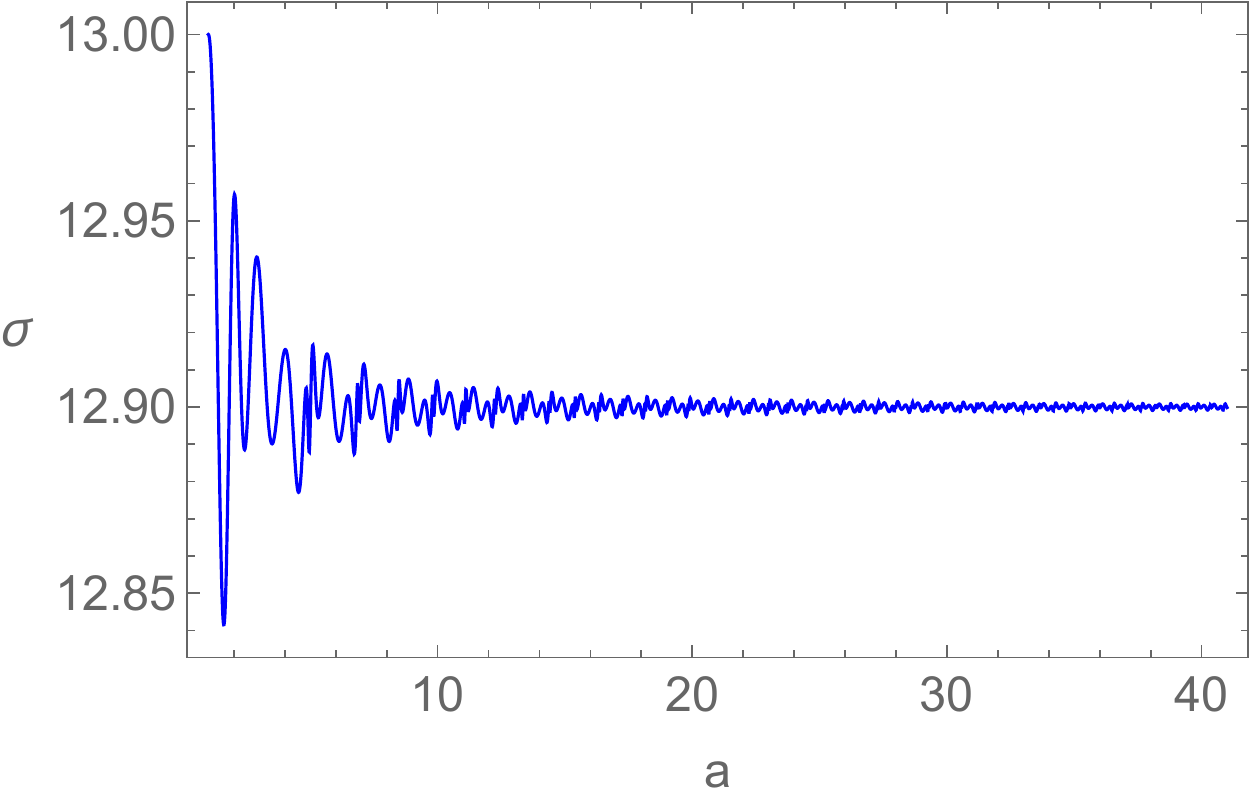} \\
    \includegraphics[scale=0.65]{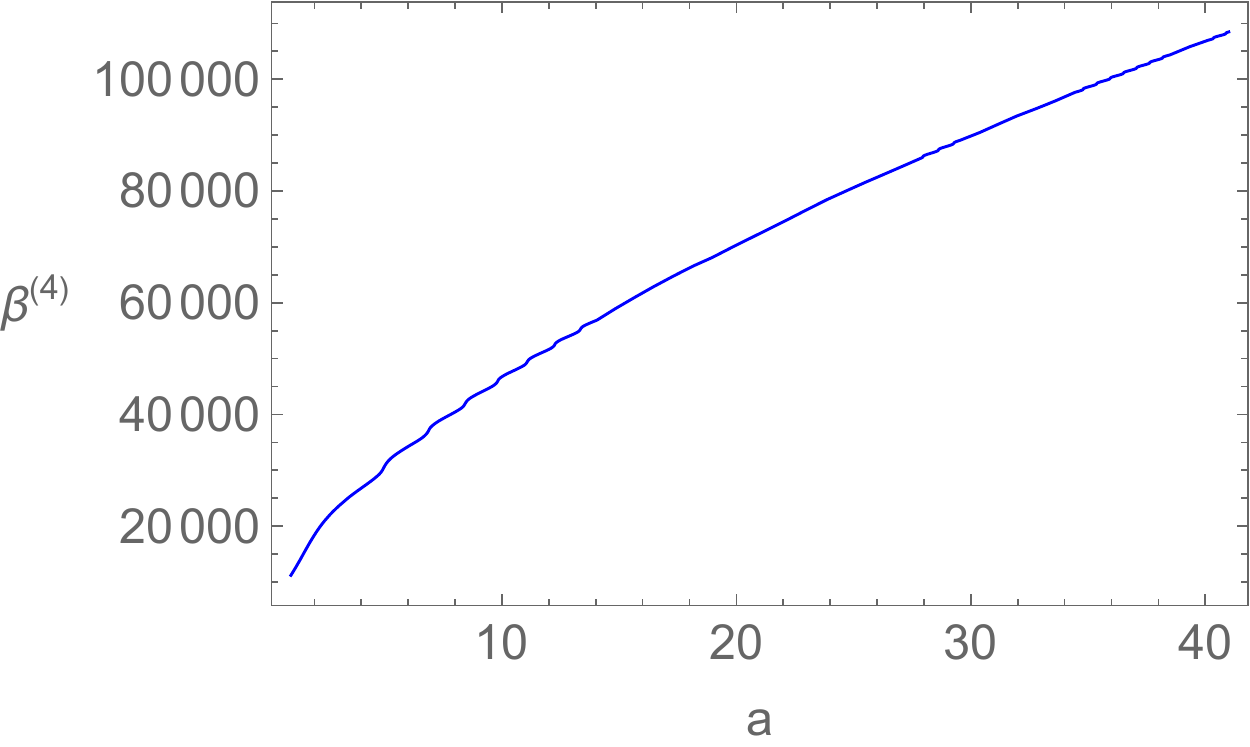} \;\;\;
    \includegraphics[scale=0.6]{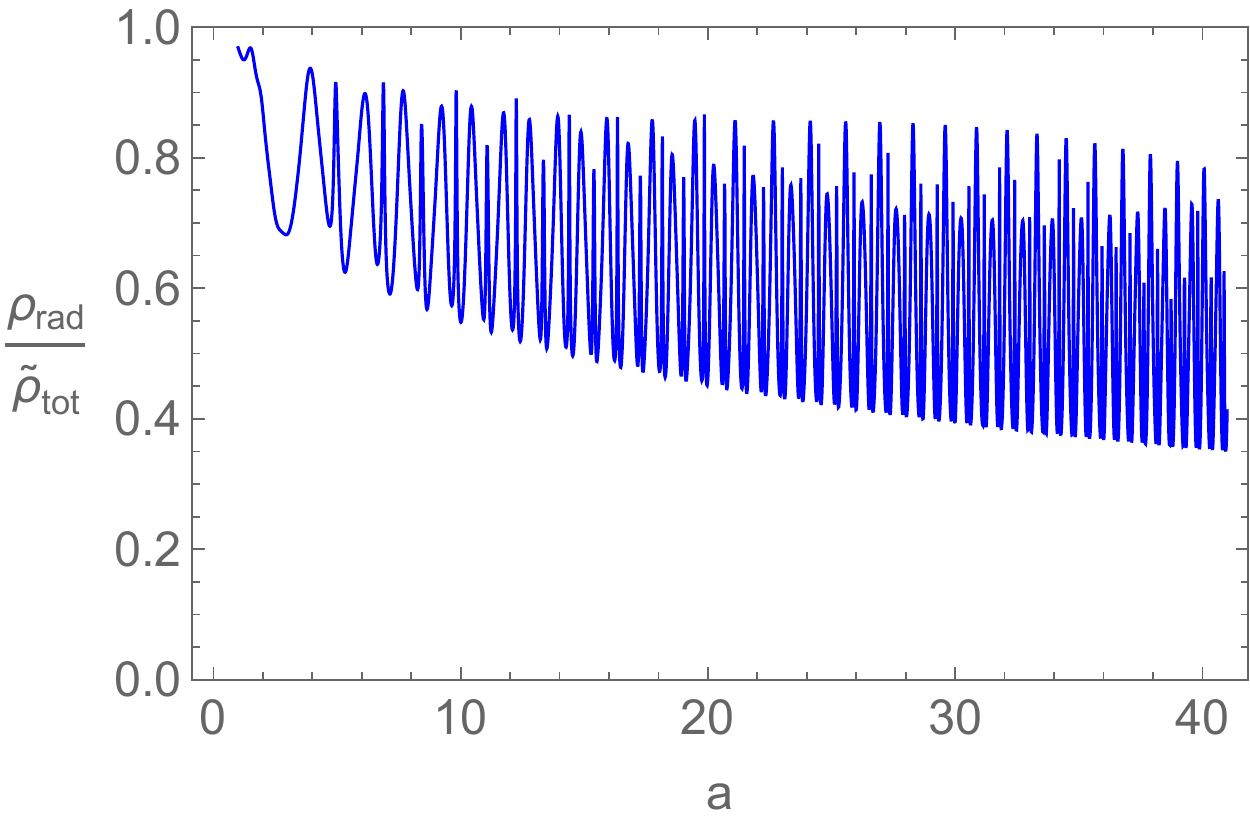} 
  \end{center}
\caption{The 3D scale factor~$a$ as a function of $t$, and the 2D scale factor~$b$, the inverse temperature measured by $M_4$, 
i.e., $\beta^{(4)}$ and the ratio~$\rho_{\rm rad}/\tilde{\rho}_{\rm tot}$ as functions of the 3D scale factor~$a$. 
The parameters are chosen as (\ref{4Dchoice:1}) and (\ref{4Dchoice:2}). 
Note that $a=41$ corresponds to $t=10^5$.}
\label{rtorhovsa}
\end{figure}

For the plots in the previous subsection, we can use the 3D scale factor~$a(t)$ instead of the time~$t$.\footnote{
Since we have chosen the initial value as $a(0)=1$, this equals $e^{N(t)}$, where $N(t)$ is the e-folding number 
for the expanding 3D space.  
}
Fig.~\ref{rtorhovsa} shows the inverse temperature measured by $M_4$, i.e., $\beta_{(4)}$, the scale factor for $S^2$, i.e., $b$, 
and the ratio~$\rho_{\rm rad}/\tilde{\rho}_{\rm tot}$ as functions of the 3D scale factor~$a$. 
The parameters are chosen as (\ref{choice:ini_values}) and $\sigma_\ast=12.9$, which correspond to
\begin{align}
 m^{(4)} &= 8.92\times 10^{-6}, \nonumber\\
 m_{\rm KK}^{(4)} &\equiv \frac{1}{R_{\rm phys}^{(4)}} = \frac{1}{2\sqrt{\pi}b_{\rm vac}^2} = \frac{2e^{-\sigma_\ast}}{\sqrt{\pi}}
 = 2.82\times 10^{-6}, 
 \label{4Dchoice:1}
\end{align}
and the initial values are chosen as (\ref{initial_values}). 
In particular, $\beta(0)=10$ corresponds to 
\begin{align}
 \beta^{(4)}(0) = 1.12\times 10^4. 
 \label{4Dchoice:2}
\end{align}
From the plots, we can see that the modulus~$b$ is stabilized in about 3 e-foldings for the 3D space expansion. 
The stabilization time scale for the other modulus~$\sigma$ is shorter than it 
when $m\gtrsim m_{\rm KK}$.

\section{Conclusions}
\label{sec:con}

We investigated the time evolutions of a higher-dimensional space during the process of the moduli stabilization, 
which is assumed to occur in the radiation dominated era. 
In many of the previous related works, 
the analyses have been performed within 4D effective field theories, e.g.,~\cite{Coughlan:1983ci,Goncharov:1984qm,Ellis:1986zt,Banks:1993en,deCarlos:1993wie,Dine:1995kz,Moroi:2001ct}. 
This means that only restricted parameter spaces, in which the moduli stabilization mass scale~$m$ is smaller than 
the compactification scale~$m_{\rm KK}$, have been investigated. 
In this paper, we have studied the spacetime evolution by solving the higher dimensional (background) field equations 
in order to extend the analysis beyond the 4D effective theory. 
As a previous research along this direction, the authors of Ref.~\cite{Maeda:1984gq} discussed 
the spacetime evolution of the Salam-Sezgin model~\cite{Salam:1984cj} in the presence of the radiation. 
However, they have set the equation of state for the radiation to a constant by hand. 
Besides, one combination of the moduli remains to be unfixed in their model. 
Here, we have treated the energy density and the pressures as independent thermodynamic quantities, 
and introduced the dilaton potential~$V_{\rm stab}(\sigma)$ in order to stabilize all the moduli.

We have checked that when the mass scale of the stabilization potential is much smaller than the compactification scale ($m\ll m_{\rm KK}$), 
the moduli oscillation around the potential minimum rapidly dominates the energy density 
over the radiation, and the non-compact 3D space expands as the matter-dominated universe~$a(t)\sim t^{2/3}$. 
This is a well known result obtained from the analysis of the 4D effective theory. 
In contrast, when $m\gtrsim m_{\rm KK}$, the situation changes. 
In this case, the 4D effective theory approach is no longer valid, and 
we have found that the energy density of the radiation~$\rho_{\rm rad}$ remains to be non-negligible 
for a long time even after the moduli have been stabilized. 
In fact, the time scale for the moduli stabilization is mainly determined by 
the mass scale of the stabilization potential~$V_{\rm stab}(\sigma)$, i.e., $t_{\rm stab}\sim 10m^{-1}$, 
while the time necessary for the moduli oscillation to dominate the energy density is determined by 
the stabilized size of the compact space~$b_{\rm vac}$. 
For the parameters that realize $m>m_{\rm KK}$, the latter can be much larger than the former. 
As a result, the 3D space expands slower than the usual matter-dominated universe. 
In such a situation, we have also checked that the equation of state for the radiation~$w_{\rm rad}$ remains smaller than 
the 3D-space value~$1/3$ for much longer time than the moduli-stabilization time scale~$t_{\rm stab}$. 
Namely, the radiation feels the extra dimensions for a while even after the moduli are fixed. 

In our model, there is an upper bound on the stabilized size of $S^2$. 
This stems from the specific structure of the stabilization potential we introduced, 
and is not a general property of the moduli stabilization. 
Hence, with an appropriate stabilization mechanism, we can realize a situation that $m$ is arbitrarily larger than $m_{\rm KK}$. 
In such a case, there is enough time for the deviation of the spacetime evolution 
from that of the conventional 4D moduli oscillation to affect the current observable. 
This can lead to some signals that suggest the existence of the extra dimensions. 
We will discuss this possibility including the generalization of Salam-Sezgin model \cite{Parameswaran:2006db,Parameswaran:2007cb,Parameswaran:2009bt} in our subsequent works.

In this work, we have focused on the effects of moduli stabilization on the spacetime evolution. 
Thus, we just added the potential~$V_{\rm stab}(\sigma)$ by hand. 
Of course, its origin has to be clarified when we construct a complete model. 
Furthermore, the moduli fields will eventually decay into matter fields, and their energy converts into the radiation. In conventional scenarios, the moduli fields dominate the energy density of the universe. 
However, our results suggest the possibility that there is no such a moduli-dominated era before their decay to the radiation. 
In order to discuss the full thermal history of the universe, we need to check whether this is the case.
We also leave these issues for the subsequent papers.

\subsection*{Acknowledgements}

  H. O. was supported in part by JSPS KAKENHI Grant Number JP20K14477.

\end{document}